\documentclass[fleqn,usenatbib]{mnras}

\usepackage{newtxtext,newtxmath}
\usepackage[T1]{fontenc}

\DeclareRobustCommand{\VAN}[3]{#2}
\let\VANthebibliography\thebibliography
\def\thebibliography{\DeclareRobustCommand{\VAN}[3]{##3}\VANthebibliography}

\usepackage{graphicx}	% Including figure files
\usepackage{amsmath}	% Advanced maths commands

\usepackage[normalem]{ulem}
\usepackage{float}
\usepackage{hyperref}
\title[]{H$\rm{\alpha}$ Reverberation Mapping from Broad-band Photometry of Dwarf Seyfert 1 Galaxy NGC 4395}

\author[Huapeng Gu]{
Huapeng Gu,$^{1,2}$
Xue-Bing Wu,$^{1,2}$\thanks{E-mail: wuxb@pku.edu.cn}
Yuhan Wen,$^{1,2}$
Qinchun Ma,$^{1,2}$
\newauthor 
{and Hengxiao Guo$^{3}$}
\\
$^{1}$Department of Astronomy, School of Physics, Peking University, Beijing 100871, China\\
$^{2}$Kavli Institute for Astronomy and Astrophysics, Peking University, Beijing 100871, China\\
$^{3}$Key Laboratory for Research in Galaxies and Cosmology, Shanghai Astronomical Observatory, Chinese Academy of Sciences, 80 Nandan Road, Shanghai
200030, China
}

\date{Accepted XXX. Received YYY; in original form ZZZ}

\pubyear{2015}

\begin{document}
\label{firstpage}
\pagerange{\pageref{firstpage}--\pageref{lastpage}}
\maketitle

\begin{abstract}
NGC 4395 is a dwarf Seyfert 1 galaxy with a possible intermediate-mass black hole of several $\rm{10^4}$ solar masses in its center. As a well-studied object, its broad line region size has been measured via H$\rm{\alpha}$ time lag in numerous spectroscopic reverberation mapping (SRM) and narrow-band photometric reverberation mapping (PRM) campaigns. Here we present its H$\rm{\alpha}$ time lag measurement using broad-band photometric data, with the application of our newly-developed ICCF-Cut method as well as the JAVELIN and $\chi ^2$ methods. utilizing the minute-cadence multi-band light curves obtained from the $\rm{2}$m FTN and $\rm{10.4}$m GTC telescopes in recent works, we measured its H$\rm{\alpha}$ lag as approximately $40 \sim 90$ minutes from broad-band PRM. With the H$\rm{\alpha}$ emission line velocity dispersion, we calculated its central black hole mass as $\rm M_{\rm BH} = (8\pm4) \times 10^3\, M_{\rm \odot}$. These results are comparable with previous results obtained by narrow-band PRM and SRM, providing further support to an intermediate-mass black hole in NGC 4395. In addition, our study also validates the ICCF-Cut as an effective method for broad-band PRM, which holds the potential for widespread application in the era of large multi-epoch, high-cadence photometric surveys.
\end{abstract}

\begin{keywords}
galaxies: active – galaxies: individual: NGC 4395 – galaxies: photometry – galaxies: Seyfert
\end{keywords}

%%%%%%%%%%%%%%%%%%%%%%%%%%%%%%%%%%%%%%%%%%%%%%%%%%

%%%%%%%%%%%%%%%%% BODY OF PAPER %%%%%%%%%%%%%%%%%%

\section{Introduction}           %% first-level sections will be auto-capitalized
\label{sect:introduction}

NGC 4395, a nearby dwarf galaxy which is only $\rm{4.4}$ Mpc away from us \citep{distance}, has been a subject of extensive research due to its unique characteristics. As a Seyfert 1 galaxy, it hosts the least luminous active galactic nucleus (AGN) known to date \citep{F1993}, whose bolometric luminosity is less than $\rm{10^{41}\, erg\, s^{-1}}$ \citep{NSC2015}. Unlike conventional AGNs, the full width at half maximum (FWHM) of the broad component of its H$\rm{\alpha}$ emission line is only about $\rm{500 \sim 1500\,km\, s^{-1}}$ \citep{Woo2019, Cho2021}, a value that is comparable to the typical width of narrow emission lines in normal AGNs. These properties make it a unique sample to study AGNs at the low luminosity end. Inspiringly, a plethora of evidence suggests that the mass of its central black hole (BH) is most likely below $\rm{10^{6}\, M_{\rm \odot}}$ \citep{F2003}, indicating the potential existence of an intermediate-mass black hole (IMBH) in the center of the galaxy \citep{evidenceIMBH1999,evidenceIMBH2003}. Since the central BH masses of most observed AGNs fall within the range of $\rm{10^{7} \sim 10^{9} \, M_{\rm \odot}}$ \citep{AGNmass,AGNmass2}, the measurement of the central BH mass of NGC 4395 is of great significance to our understanding of the structure and formation of such low luminosity AGNs, as well as the potential presence of IMBHs in galaxy centers.

Given the inherent challenges associated with probing such BHs via spatially resolution techniques, currently the most feasible approach to measuring AGN central BH mass is the reverberation mapping (RM) \citep{RM}. This technique provides us with an opportunity to measure the BH mass by determining the size of the broad-line region (BLR) \citep{unifiedmodel} through long-term monitoring rather than spatially-resolved observations. The distance of the BLR from the central BH results in the variation of the broad emission line lagging behind the variation of the continuum radiation from the accretion disk. The time lag ${t}$ between the light curves of the continuum and emission line can be measured, which just corresponds to the BLR size, i.e. ${R_{\rm BLR}=ct}$, where ${c}$ is the speed of light. Combined with the velocity dispersion ${\sigma_{\rm disp}}$ of the broad emission line measured from the spectrum, we can estimate the central BH mass kinematically using the equation below: 
\begin{equation}\label{eq1}
  {M_{\rm BH}}= \textit{f} \frac{R_{\rm BLR}\sigma_{\rm disp}^2}{G} ,
\end{equation}
\noindent where ${G}$ is the gravitational constant and $f$ is a dimensionless factor related to the geometry of the BLR \citep{Woo2015}.

Observations of NGC 4395 have revealed short-timescale variability in its luminosity across different wavebands \citep{4395variation,4395variation2,4395variation3,4395variation4,4395variation5}. Several RM campaigns have been conducted, focusing on the measurement of the central BH mass of NGC 4395 through its H$\rm{\alpha}$ emission line. Based on broad-band photometry, \citet{CC-AC2012} computed the H$\rm{\alpha}$ lag as $3.6\pm0.8$ hours by employing the CCF-ACF method. In a separate study, \citet{Woo2019} utilized narrow-band photometry to acquire the H$\rm{\alpha}$ light curve. They reported the time delay of H$\rm{\alpha}$ as $\rm {83 \pm 14}$ min. Combining this with the H$\rm{\alpha}$ velocity dispersion ${\sigma_{\rm disp}}=\rm{426\,km\, s^{-1}}$, they obtained its BH mass as $\rm{(9.1 \pm 1.6)\times 10^3\,M_{\rm \odot}} $. As a follow-up, \citet{Cho2021} carried out high-cadence spectroscopic monitoring on NGC 4395 and calculated its H$\rm{\alpha}$ lag, which is less than 3 hours. With the measured FWHM of broad H$\rm{\alpha}$ as $\rm{586\,km\, s^{-1}}$, they updated its BH mass up to $\rm{(1.7 \pm 0.3)\times 10^4\,M_{\rm \odot}} $. Apart from the results of the H$\rm{\alpha}$ RM campaigns, an earlier RM based on its C IV broad emission line suggested a higher BH mass of several $\rm{10^5\,M_{\rm \odot}} $\citep{C4RM}. Some alternative methods have also been employed. For instance, the break in power density spectrum (PDS) yielded the result of $\rm{10^4 \sim 10^5 \,M_{\rm \odot}}$ \citep{evidenceIMBH2003}. Gas dynamical modeling \citep{NSC2015, 4395spec} and the method based on spectroscopic SE relationship \citep{SErelation} have also been utilized, both reporting a BH mass of several $\rm{10^5 \,M_{\rm \odot}}$ . The continuum RM technique has also been applied, resulting in a BH mass estimation of several $\rm{10^4 \,M_{\rm \odot}}$ \citep{2023ApJ...948L..23W}. The exact mass of the central BH in NGC 4395 remains a subject of ongoing debate. This underscores the need for more precise observations to obtain a more reliable result.

The critical aspect of H$\rm{\alpha}$ RM is the accurate determination of the flux of the broad H$\rm{\alpha}$ emission line. Some previous photometric RM (PRM) works have employed specific narrow bands to trace the emission lines \citep{2011A&A...535A..73H,2012A&A...545A..84P}. However, narrow-band photometry presents certain limitations. The band width is limited, necessitating an extension of the exposure time to acquire high signal-to-noise ratio (SNR) H$\rm{\alpha}$ light curves, let alone spectroscopic observations. As a result, the time resolution of the light curves is compromised when using narrow-band photometry or spectroscopic observations. This could result in significant uncertainties in the calculation of the time lag, as the variability timescale of NGC 4395 is only about hours \citep{M,McHardy+etal+2023}. To achieve a more precise estimation of the H$\rm{\alpha}$ time lag, the observing cadence must be on the order of minutes. At present, the most effective approach to acquiring high-cadence light curves for RM is through broad-band photometry. However, the broad-band photometric data, although more precise in both time and flux, has a crucial weakness. The continuum component is dominant, meaning that the H$\rm{\alpha}$ broad emission line contributes only a small fraction of the total flux in the broad-band. Compared with the narrow-band PRM, the broad-band PRM is more sensitive to the errors in the determined continuum component, which can significantly influence the extracted emission line light curve. Therefore, the precise extraction of the H$\rm{\alpha}$ line component from the broad-band presents a formidable challenge.

Previously, the broad-band PRM is based on the ``CCF-ACF" method \citep{2012ApJ...747...62C}. Using the multi-band photometric data, this method empirically subtracts the continuum contribution from the broad-band. However, it operates under the assumption of a fixed ratio for the continuum flux, while this ratio may exhibit variations among different AGNs. In our previous work \citep{Ma2023}, we improved this method by incorporating a single-epoch spectrum to determine the continuum flux ratio. It has been demonstrated that our method, namely ICCF-Cut, yields successful results with some high-quality broad-band photometric data and provides the H$\rm{\alpha}$ time lag with respect to the continuum band (see \citet{Ma2023}). In this work, we apply this method to NGC 4395. \citet{M} carried out a minute-cadence broad-band $ugriz$ photometric observation on NGC 4395 using the $\rm{2}$m FTN telescope for two successive nights, with the aim of analyzing its accretion disk size. Since the r-band encompasses the H$\rm{\alpha}$ emission line, the ICCF-Cut method can be employed to extract the H$\rm{\alpha}$ light curve and determine its time delay with respect to the g-band. In addition, \citet{McHardy+etal+2023} also conducted simultaneous $ugriz$ photometry of NGC 4395 using the $\rm{10.4}$m GTC telescope. In this work, we measure the broad H$\rm{\alpha}$ time lag from these photometric data. Moreover, \citet{Ma2023,2024arXiv240310223M} have demonstrated that the combination of ICCF-Cut, JAVELIN Pmap model (\citealt{Zu2011,Zu2013,Zu2016}, see also in Sec.~\ref{2.2}) and $\rm{\chi^2}$ method (\citealt{Czerny2013,Bao2022}, see also in Sec.~\ref{2.3}) can be an effective approach for broad-band photometric reverberation mapping. Therefore, to enhance the reliability of our result, we also apply the JAVELIN Pmap model method and the $\rm{\chi^2}$ method to calculate the time lag of H$\rm{\alpha}$.

As explored by previous works \citep{2013ApJ...769..124C,2014ApJ...785..140C,2023A&A...675A.163C}, the broad-band PRM, with the usage of the ``CCF-ACF" method, has the potential for widespread application in the era of large multi-epoch, high-cadence photometric surveys such as LSST (Legacy Survey of Space and Time, \citealt{2017arXiv170804058L}). Now with the advent of our ICCF-Cut method, the continuum flux ratio of the AGNs can be determined individually, resulting in better measurements of the BLR size. The JAVELIN Pmap model and the $\rm{\chi^2}$ method can be applied as well. The combination of the three methods allows for cross-validation of the results. Therefore, our methods could be extensively implemented in the future. Moreover, in these surveys, it has been predicted that BH masses could be estimated via continuum RM \citep{2023ApJ...948L..23W}, and our methods are anticipated to have better validation.

This paper is arranged as follows: in Sec.~\ref{sect:method} we introduce the ICCF-Cut, JAVELIN and $\rm{\chi^2}$ methods for calculating the H$\rm{\alpha}$ time lag. In Sec.~\ref{sect:data processing}, we discuss the processing of the spectrum and the photometric data. Sec.~\ref{sect:result} presents the results, including the obtained H$\rm{\alpha}$ time lag via the three methods and the central BH mass estimation of NGC 4395. Some discussions on the results are given in Sec.~\ref{sect:discussion}, and finally the summary is given in Sec.~\ref{sect:conclusion}.

\section{Method}
\label{sect:method}

In our analysis, we employ three different methods: the ICCF-Cut, JAVELIN, and $\rm{\chi^2}$ methods to calculate the time lag between the H$\rm{\alpha}$ and g-band light curves. This section provides a briefly introduction to the fundamental principles of three methods.

\subsection{ICCF-Cut Method\label{2.1}}

The ICCF (Interpolation $\&$ Cross-Correlation Function, \citet{1987ApJS...65....1G,1994PASP..106..879W,Peterson1998}) method is a well-established technique for computing the time lag between two light curves, particularly in the context of spectroscopic RM (SRM) and continuum PRM. The time lag is obtained by calculating the cross-correlation function of two light curves. However, it is not directly applicable to the case where one photometric band is a mix of continuum radiation and emission lines. In the previous work \citep{Ma2023}, we proposed the ``ICCF-Cut" method. This approach involves deducting the continuum's contribution to isolate the line component out of the line band, utilizing a single-epoch spectrum. Subsequently, the ICCF method is applied to calculate its lag relative to the continuum band (we use ``continuum band" to refer to the band with a negligible emission line component, and ``line band" to refer to the band with significant emission line contribution). In the following we will only briefly introduce its principle. More details can be found in \citet{Ma2023} and \citet{2024arXiv240310223M} for a comprehensive understanding of this method. Also, the code for the ICCF-Cut method is available at \href{https://github.com/PhotoRM/ICCF-Cut} {https://github.com/PhotoRM/ICCF-Cut}.

In the specific case of NGC 4395, the r-band contains the broad H$\rm{\alpha}$ emission line from the BLR and the continuum radiation emanating from the accretion disk. The g-band can be approximately regarded as the continuum band comprising solely of flux originating from the accretion disk\footnote{The contribution of emission lines such as H$\rm{\beta}$ is very small in the g-band. This will not affect the measurement of H$\rm{\alpha}$ lag. See the discussion in \citet{Ma2023}}. To estimate the contribution of broad H$\rm{\alpha}$ emission line in the r-band, we utilize a single-epoch spectrum of r-band, as depicted in Figure~\ref{Fig1}. From this, we derive the flux ratio of the broad H$\rm{\alpha}$ line in the r-band, represented by $k$. The continuum flux in the r-band can be inferred from the g-band light curve, since they both originate from the accretion disk. Subsequently, the following equation can be employed to derive the H$\rm{\alpha}$ light curve:
\begin{equation}\label{eq2}
  {F_{\rm H \alpha}(t)=F_{\rm r}(t)-\alpha F_{\rm g}(t)\frac{\overline{F_{\rm r}}}{\overline{F_{\rm g}}}},
\end{equation}
\noindent where ${\overline{F_{\rm g}}}$ and ${\overline{F_{\rm r}}}$ denote the average flux of the g and r-band light curves respectively, and ${\alpha=1-}\textit{k}$ is the continuum fraction in r-band. The term $ F_{\rm g}(t) / \overline{F_{\rm g}}$ reflects the variation of the continuum, thereby ${\alpha F_{\rm g}(t) \overline{F_{\rm r}} / \overline{F_{\rm g}}}$ is just the flux of the continuum in the r-band. Note that this equation may over-subtract the continuum component (see \citealt{Ma2023}). A more appropriate approach is to subtract the smallest possible continuum comnonent and retaining as much of the H$\rm \alpha$ contribution as possible, i.e:
\begin{equation}\label{eq3}
  {F_{\rm H \alpha}(t)=F_{\rm r}(t)-\alpha F_{\rm g}(t) \cdot \min _\tau\left(\frac{F_{\rm r}(\tau)}{F_{\rm g}(\tau)}\right)}.
\end{equation}
\noindent Compared with the ``CCF-ACF" method \citep{2012ApJ...747...62C}, our method provides a more realistic approach to extracting the broad H$\rm{\alpha}$ components. Subsequently, the ICCF method can be applied to calculate the time lag between this extracted H$\rm{\alpha}$ light curve and the continuum band light curve.

There are two things to note here. Firstly, while there is a potential time lag between the continuum components in the g and r-bands due to the physical size of the accretion disk, this lag is typically negligible compared to the lag of the broad emission lines. This aspect is further discussed in Sec.~\ref{3.2} and ~\ref{5.1}. Secondly, unlike ICCF, the ICCF-Cut method relies on the absolute flux of the two bands, not merely the light curve profiles according to Eq.~(\ref{eq3}). Consequently, it is necessary to take into account the emissions arising from the narrow-line region (NLR) and the host galaxy, which remain relatively constant over time. The deduction of the narrow emission lines will be addressed in Sec.~\ref{3.1}. The host galaxy contribution is quite complicated. Unlike typical AGNs, NGC 4395 harbors a nuclear star cluster (NSC) \citep{F2003,Carson2015,NSC2015}, which may contribute more significantly to the flux than the host galaxy itself. Determining their precise contributions from the light curves presents a challenge. However, our findings indicate that the host galaxy and the NSC do not significantly impact the measurement of the H$\rm{\alpha}$ time lag. Therefore, we initially disregard their contributions, with a more in-depth discussion on their potential influence provided in Sec.~\ref{5.2}. 

\subsection{JAVELIN Method\label{2.2}}
JAVELIN (The Just Another Vehicle for Estimating Lags In Nuclei, \citep{Zu2011,Zu2013,Zu2016}) is another method to calculate the time lag between two light curves. It is based on the Damped Random Walk (DRW) model \citep{Kelly2009}, which employs a set of parameters, including the time lag, to fit the two light curves. Afterwards, the distribution of parameters is given through the Markov Chain Monte Carlo (MCMC) method. This approach has seen extensive application in the past RM campaigns.

JAVELIN offers several models tailored for light curve fitting. The Rmap model, which treats the two input light curves as representing distinct components--one as the pure continuum and the other as the pure emission line, is currently well-regarded. Nevertheless, this model is not suited for broad-band photometry. Consequently, we apply JAVELIN's Pmap model, which assumes that the line band light curve comprises a continuum component and a delayed line component. The relative amplitude of the two components can be deduced from the fitting results, which is indicative of the emission line's flux ratio within the broad-band. By comparing this ratio with that derived from spectral data, we can evaluate the reliability of JAVELIN outcome.

\subsection{$\rm{\chi^2}$ Method\label{2.3}}

The $\rm{\chi^2}$ method, as detailed in  \citet{Czerny2013,Bao2022}, employs uncertainties as weights for the data points in the light curves, which is believed to work better than ICCF for AGNs that exhibit red-noise variability. Using the extracted emission line light curve from ICCF-Cut, an alternative procedure is implemented to calculate its lag corresponding to the continuum light curve. Upon interpolating the light curves, we proceed to shift the continuum light curve by a predetermined lag. Then the $\rm{\chi^2}$ between the two light curves is calculated by
\begin{equation}\label{eq4}
  {\chi^{2}=\frac{1}{N} \sum_{i=1}^{n} \frac{\left(x_{\rm i}-A_{\rm \chi^{2} y_{\rm i}}\right)^{2}}{\delta x_{\rm i}^{2}+A_{\rm \chi^{2}}^{2} \delta y_{\rm i}^{2}}},
\end{equation}
\noindent where ${x_{\rm i}}$, ${y_i}$ are the continuum and emission line light curves, and ${\delta x_{\rm i}}$, ${\delta y_i}$ are their uncertainties. ${A_{\rm \chi^{2}}}$ is the normalized factor:
\begin{equation}\label{eq5}
  {A_{\rm \chi^{2}}= \frac{S_{\rm x y}+\left(S_{\rm x y}^{2}+4 S_{\rm x 3 y} S_{\rm x y 3}\right)^{1 / 2}}{2 S_{\rm x y 3}}},
\end{equation}
\noindent where
\begin{equation}\label{eq6}
 \begin{aligned} S_{\rm x y} &  =\sum_{\rm i=1}^{N}\left(x_{\rm i}^{2} \delta y_{\rm i}^{2}-y_{\rm i}^{2} \delta x_{\rm i}^{2}\right), \\  S_{\rm x y 3} & =\sum_{i=1}^{N} x_{\rm i} y_{\rm i} \delta y_{\rm i}^{2}, \\  S_{\rm x 3 y} & =\sum_{i=1}^{N} x_{\rm i} y_{\rm i} \delta x_{\rm i}^{2}. \end{aligned}
\end{equation}
This formula is applied iteratively with different values of the time lag. The particular lag that results in the lowest $\rm{\chi^2}$ value indicates the best match between the light curves. By employing all three aforementioned methods to calculate the time lag of the broad H$\rm{\alpha}$ emission line in NGC 4395, we can cross-verify the results. This comprehensive approach will significantly enhance the reliability of our findings.

\section{Data Processing}
\label{sect:data processing}
\subsection{Spectroscopic Data\label{3.1}}
\begin{figure}
   \centering
   \includegraphics[width=8.5cm, angle=0]{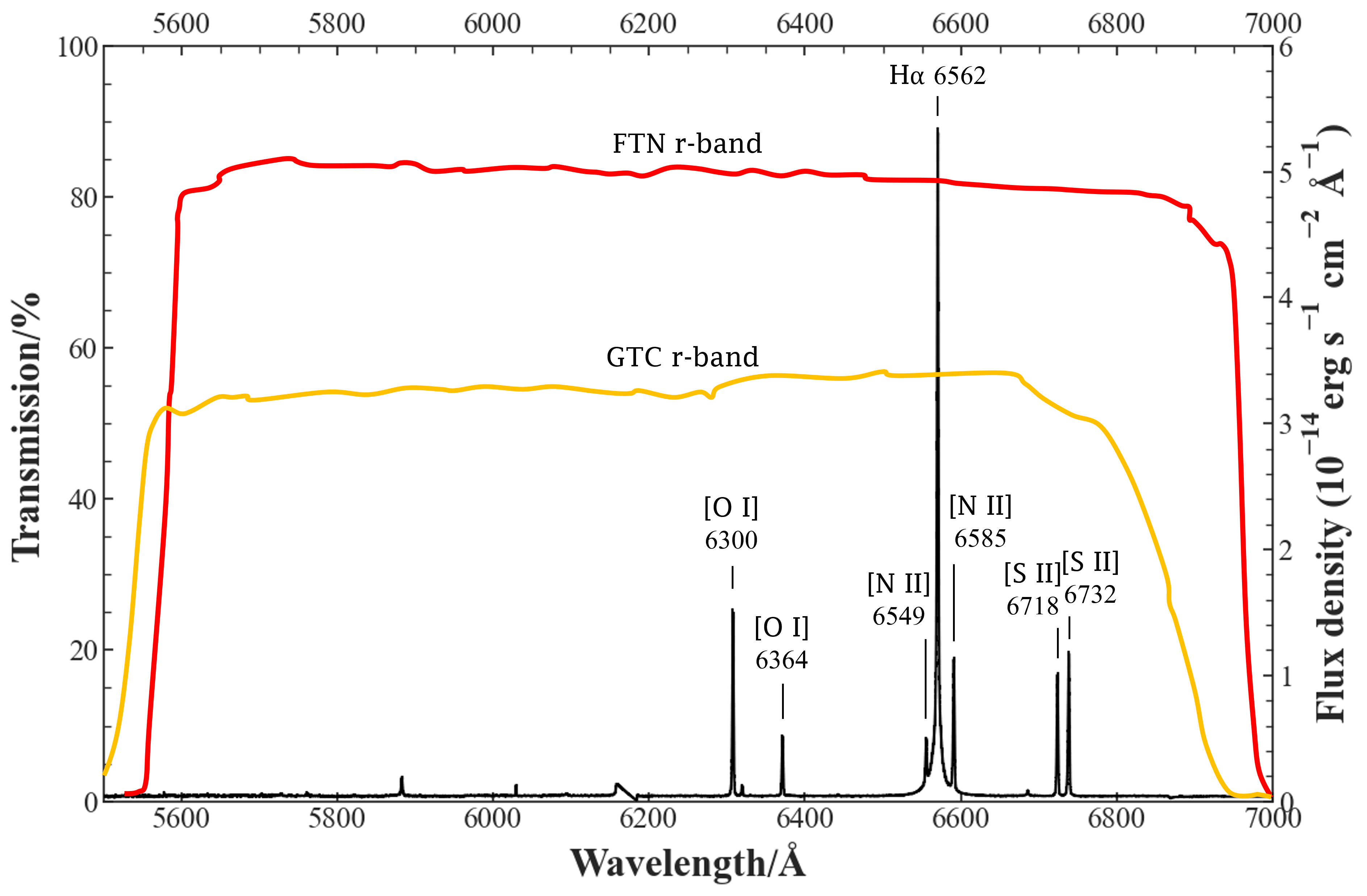}
   \caption{\small The r-band spectrum of NGC 4395 from Gemini \citep{Cho2021}. The main emission lines are annotated in this figure. The red curve represents the transmission function of the r-band for the MuSCAT3 camera \citep{MuSCAT3} on FTN, and the orange curve is the transmission function of the r-band for HIPERCAM camera \citep{HiPERCAM} on GTC. The photometric data from these telescopes are used for our PRM (see in Sec.~\ref{3.2}). The marked wavelengths of the emission lines are in the rest frame.}
   \label{Fig1}
   \end{figure}

In this section we first analyse the spectrum of NGC 4395 to obtain the properties of its emission lines. We utilize the r-band spectrum provided by \citet{Cho2021}, which was obtained from Gemini GMOS in March, 2019. The spectrum is shown in Figure~\ref{Fig1}.

\begin{figure}
   \centering
   \includegraphics[width=8.5cm, angle=0]{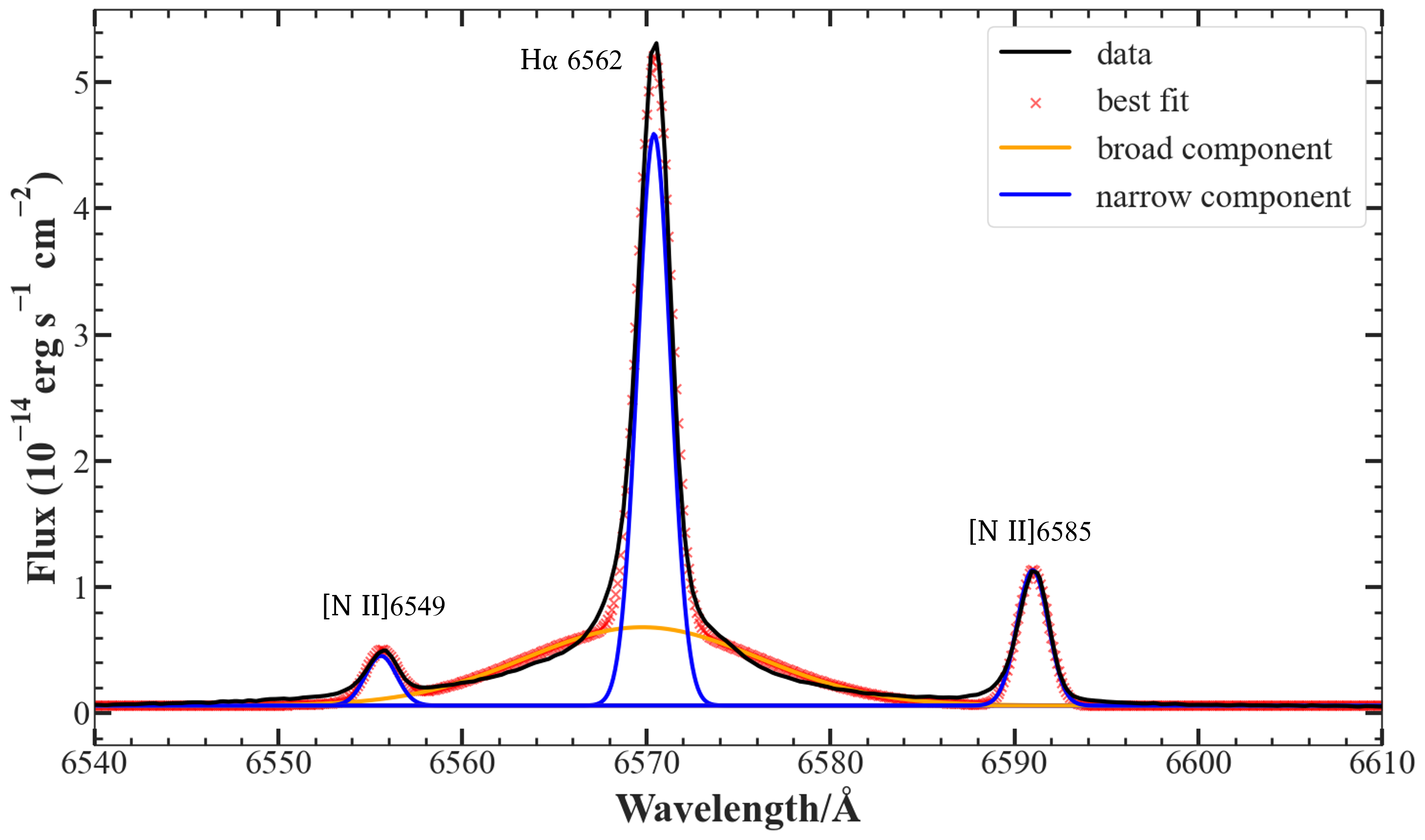}
   \includegraphics[width=85mm]{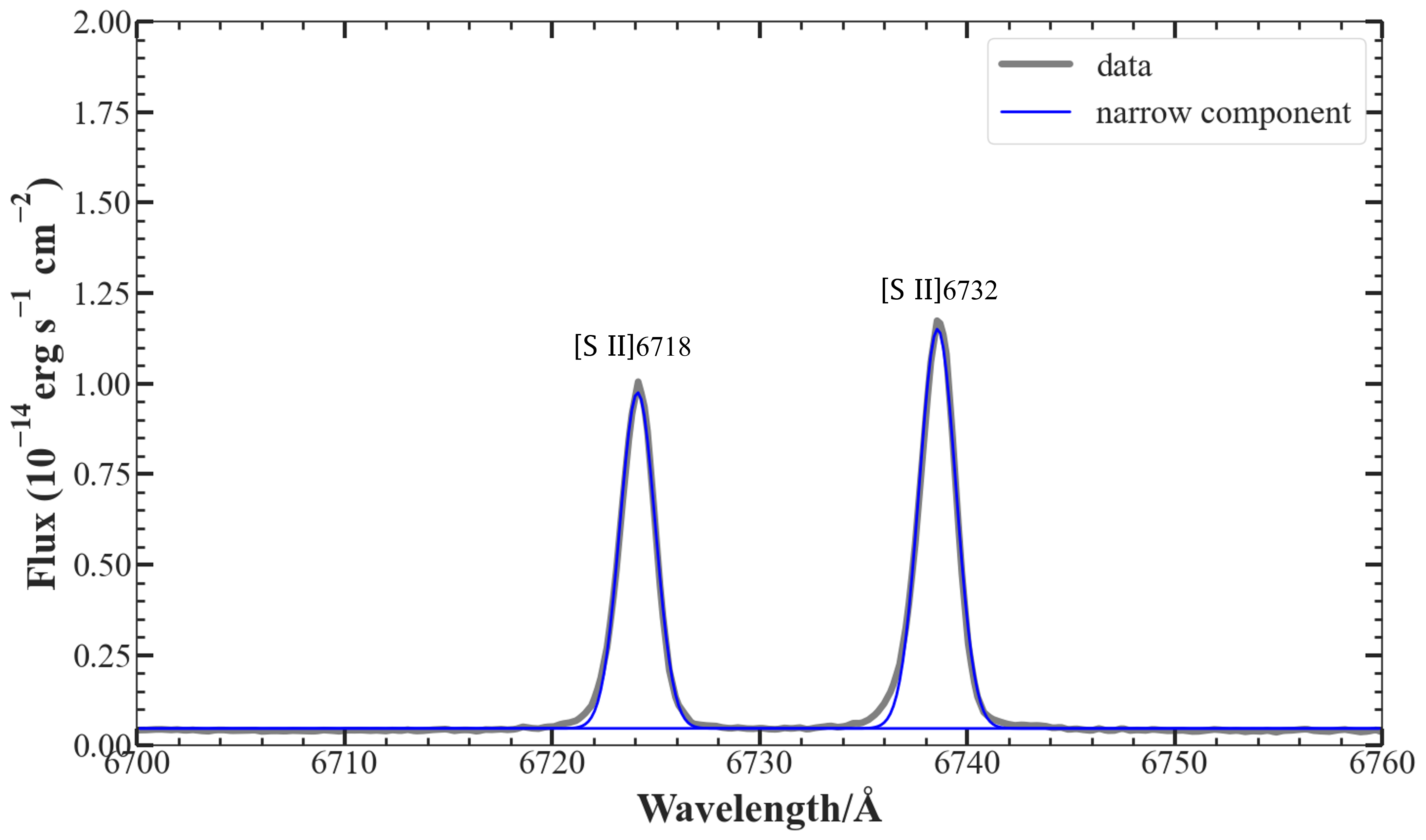}
   \includegraphics[width=85mm]{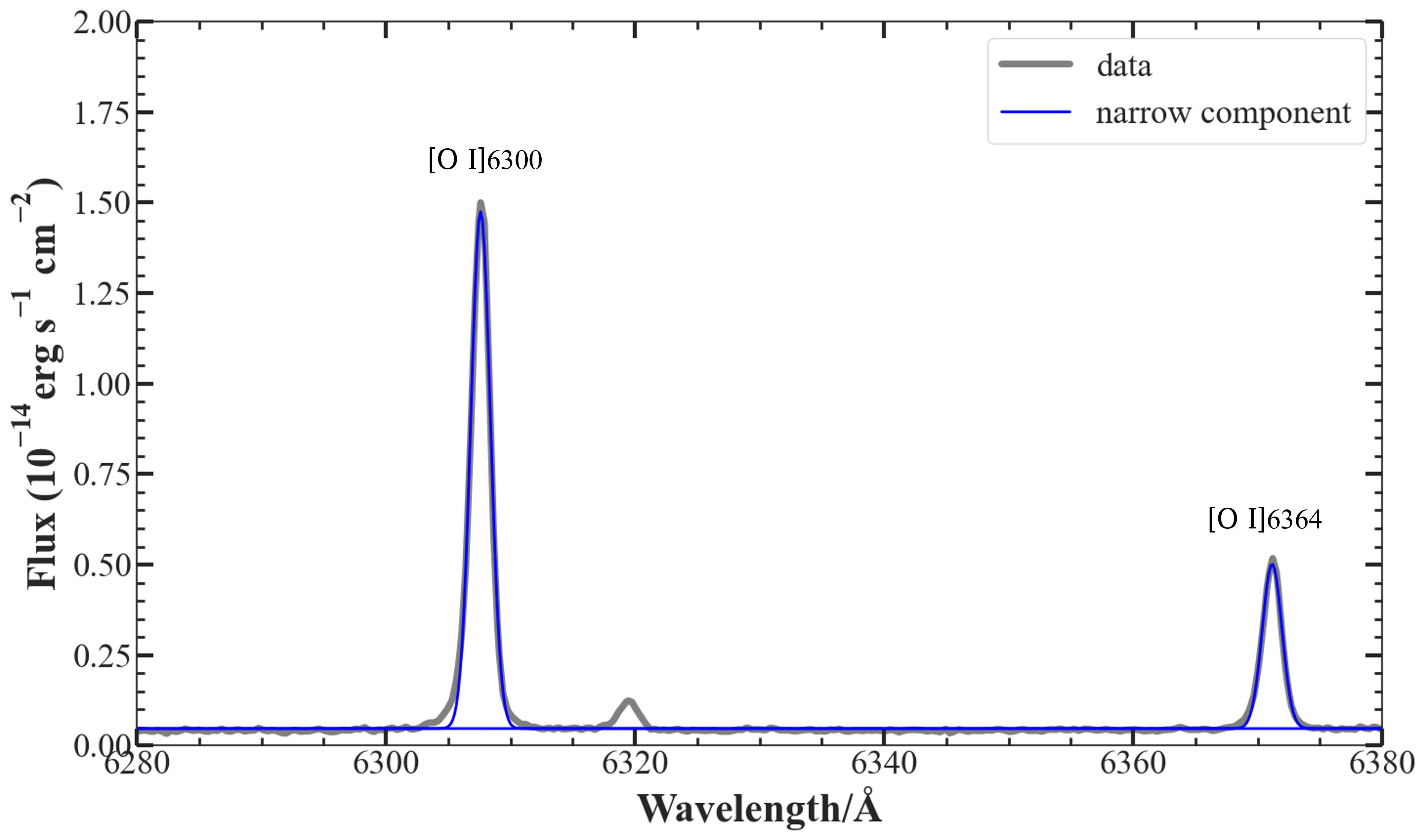}
   \includegraphics[width=8.5cm, angle=0]{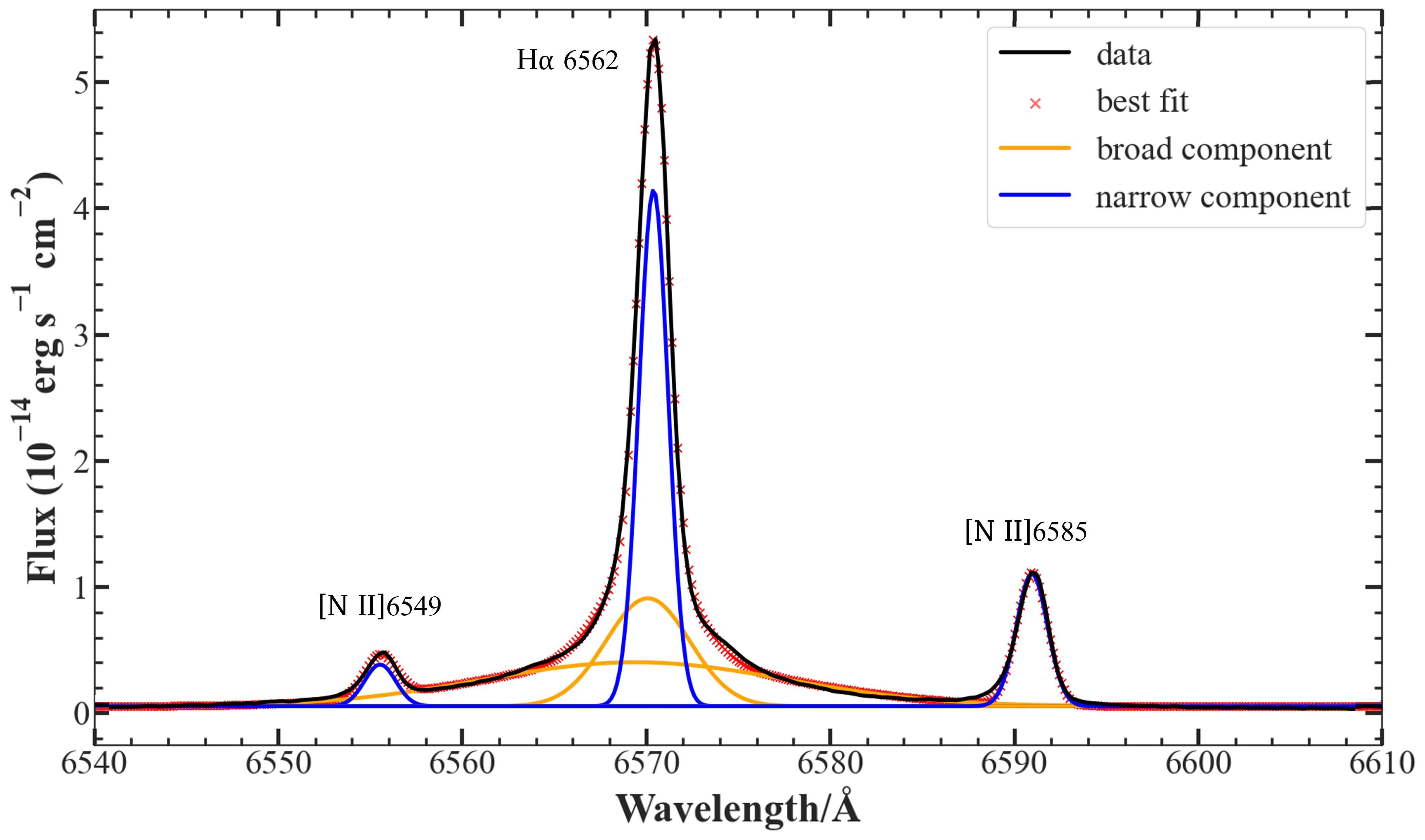}
   \caption{\small The Gaussian fitting applied to the r-band emission lines of NGC 4395. The upper panel shows the fitting for the broad and narrow components of H$\rm{\alpha}$ and two narrow {[N II]} lines. Here H$\rm{\alpha}$ consists of one narrow component (shown by the blue line) and one broad component (shown by the orange line). The black line denotes the observed data and the red crosses represent the superposition of these line components. The subsequent two panels display the fittings for two narrow {[S II]} lines and two narrow {[O I]} lines. The lowermost panel is the fitting for the H$\rm{\alpha}$ using a double-Gaussian model for the broad component. The reduced $\chi^2$ for the four fittings are $19.1$, $3.3$, $4.9$ and $7.1$ respectively. The marked wavelengths of the emission lines are in the rest frame.}
   \label{Fig2}
   \end{figure}

In this analysis, we meticulously evaluate the flux of each spectral component. We focus on three primary components: the continuum, the broad emission lines, and the narrow emission lines. The main emission lines in the r-band include two [N II] lines, two [S II] lines, two [O I] lines and the H$\rm{\alpha}$ line. H$\rm{\alpha}$ is the dominant component of the emission lines, with other emission lines being significantly weaker. Our aim is to determine the total flux for each of the three components. Intuitively, we assume that H$\rm{\alpha}$ line is composed of a broad component and a narrow component, whereas the remaining emission lines only consist of one narrow component. All components are modeled using Gaussian profiles. Figure~\ref{Fig2} shows the optimal fit for these emission lines. The fitting results, including the flux and width of these components, are enumerated in Table~\ref{Tab1}. The uncertainty of these parameters is derived through 50 Monte Carlo processes, each of which randomly selects a spectrum within the error margin of the original data and fits the parameters accordingly. The variance of the parameters across the 50 processes is considered as the error margin for these parameters. Our results indicate that the broad H$\rm{\alpha}$ component typically exhibits a width of $740$ $\rm{km \ s^{-1}}$, and the narrow lines have a typical width of $70\sim100$ $\rm{km \ s^{-1}}$.

\begin{table}
\begin{center}
\caption[]{Flux and Width (FWHM) of the Components in NGC 4395 (r-band)}\label{Tab1}

\begin{tabular}{ccc}
\hline
Component           & Flux($ 10^{-14} $ $ \rm erg \ s^{-1} \ cm^{-2}$) & FWHM($\rm{km \ s^{-1}}$) \\ \hline
Continuum           & $60.4\pm0.5$                           & -          \\ 
H$\rm{\alpha}$ broad & $11.93\pm0.05$                            & $740\pm3$        \\ 
H$\rm{\alpha}$ narrow    & $9.44\pm0.02$                            & $98.4\pm0.3$         \\ 
$\rm [N\ II]6549$    & $0.78\pm0.01$                           & $91.4\pm0.2$        \\ 
$\rm [N\ II]6585$   & $2.10\pm0.01$                           & $90.9\pm0.2$        \\ 
$\rm [O\ I]6300$          & $2.13\pm0.02$                           & $78.0\pm0.1$         \\ 
$\rm [O\ I]6364$             & $0.72\pm0.01$                           & $78.3\pm0.1$         \\  
$\rm [S\ II]6718$    & $1.60\pm0.01$                           & $81.1\pm0.2$         \\ 
$\rm [S\ II]6732$   & $1.85\pm0.01$                           & $81.3\pm0.2$         \\

\noalign{\smallskip}\hline
\end{tabular}
\end{center}
\end{table}

From the results of our spectral fitting, we are able to derive the flux ratios of broad H$\rm{\alpha}$ line and narrow emission lines in the r-band photometric data. According to Table~\ref{Tab1}, the total flux ratio of the broad H$\rm{\alpha}$ line in r-band is $14\%$, and the narrow emission lines account for $18\%$. Furthermore, to facilitate the calculation of the BH mass, the velocity dispersion of broad H$\rm{\alpha}$ line is also required. Given that our analysis is predicated on a single-epoch spectrum, we have examined the FWHM of the broad H$\rm{\alpha}$ emission line as a proxy for its velocity dispersion. The above results implies that the FWHM of the broad H$\rm{\alpha}$ emission line is $740\pm3$ $\rm{km \ s^{-1}}$. Although this value is not typical for the broad emission lines of an AGN, we still regard it as the broad component due to the diminutive size of NGC 4395, which implies a correspondingly smaller velocity dispersion. It is noteworthy that this value is smaller than $1000$ $\rm{km \ s^{-1}}$ from \citet{Woo2019} and $1500$ $\rm{km \ s^{-1}}$ reported by \citet{evidenceIMBH1999}. We also calculate the $\rm{5100 \mathring{A}}$ luminosity from this spectrum, which aligns with the value reported by \citet{evidenceIMBH1999}. The variance of the broad H$\rm{\alpha}$ line width across different epochs can affect the determination of the BH mass. 

We noticed that in \citet{Cho2021}, they incorporated additional broad components for the H$\rm{\alpha}$ in fitting this spectrum. Their modelling results (depicted in Fig. 4 of their paper) revealed two distinct broad components, with FWHMs being $\sim$ $200$ $\rm{km \ s^{-1}}$ and $\sim$ $800$ $\rm{km \ s^{-1}}$ separately. On average, they obtain the FWHM of broad H$\rm{\alpha}$ as $\sim$ $600$ $\rm{km \ s^{-1}}$, which is slightly smaller than the results from our single-Gaussian modelling for the broad H$\rm{\alpha}$. We also applied a double-Gaussian model to fit the broad component of H$\rm{\alpha}$, obtaining a similar result with smaller reduced $\chi^2$ (about $7.1$). The fitting result using the double-Gaussian model is also shown in Figure~\ref{Fig2}\footnote{Although the reduced $\chi^2$ is smaller, adopting a more complicated line profile remains controversial. Various factors such as the outflows, could cause deviations from a perfect Gaussian profile, leaving the nature of the second component uncertain. Also, in principle the line profile is not necessarily to be best described by a single-Gaussian \citep{2002ApJ...566L..71S}. Taking these into considerations, we present the fitting results using both the single-Gaussian and the double-Gaussian model.}. The two broad components have FWHMs of $240$ $\rm{km \ s^{-1}}$ and $860$ $\rm{km \ s^{-1}}$ respectively, and the average FWHM is $640$ $\rm{km \ s^{-1}}$. Since this model considers an additional component for broad H$\rm{\alpha}$, it suggests a slightly higher contribution of broad H$\rm{\alpha}$ to the r-band flux (about $18\%$). This subtle difference in broad H$\rm{\alpha}$ line flux ratio does not significantly impact the calculation of the H$\rm{\alpha}$ lag. According to our PRM method, the time lag measurement cannot be separated into two distinct components, so that we proceed with the results from the single-Gaussian model for subsequent calculations. Moreover, we note that the primary uncertainty in BH mass estimation lies in the geometry factor ``$f$" in Eq.~(\ref{eq1}), which accounts for errors arising from diverse interpretations of the BLR structure.

\subsection{Photometric Data\label{3.2}}
\begin{figure}
\centering
   \includegraphics[width=60mm]{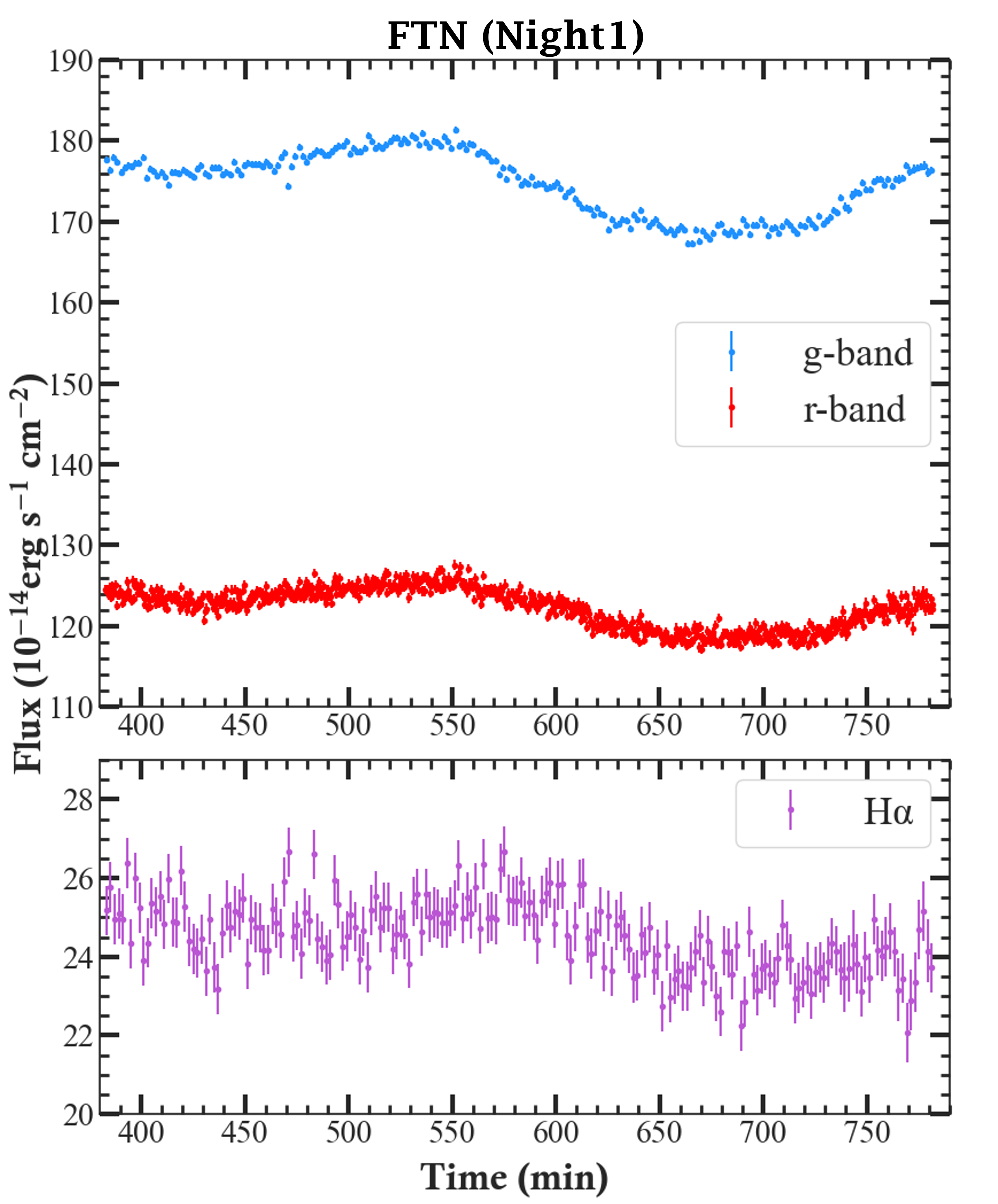}
   \includegraphics[width=60mm]{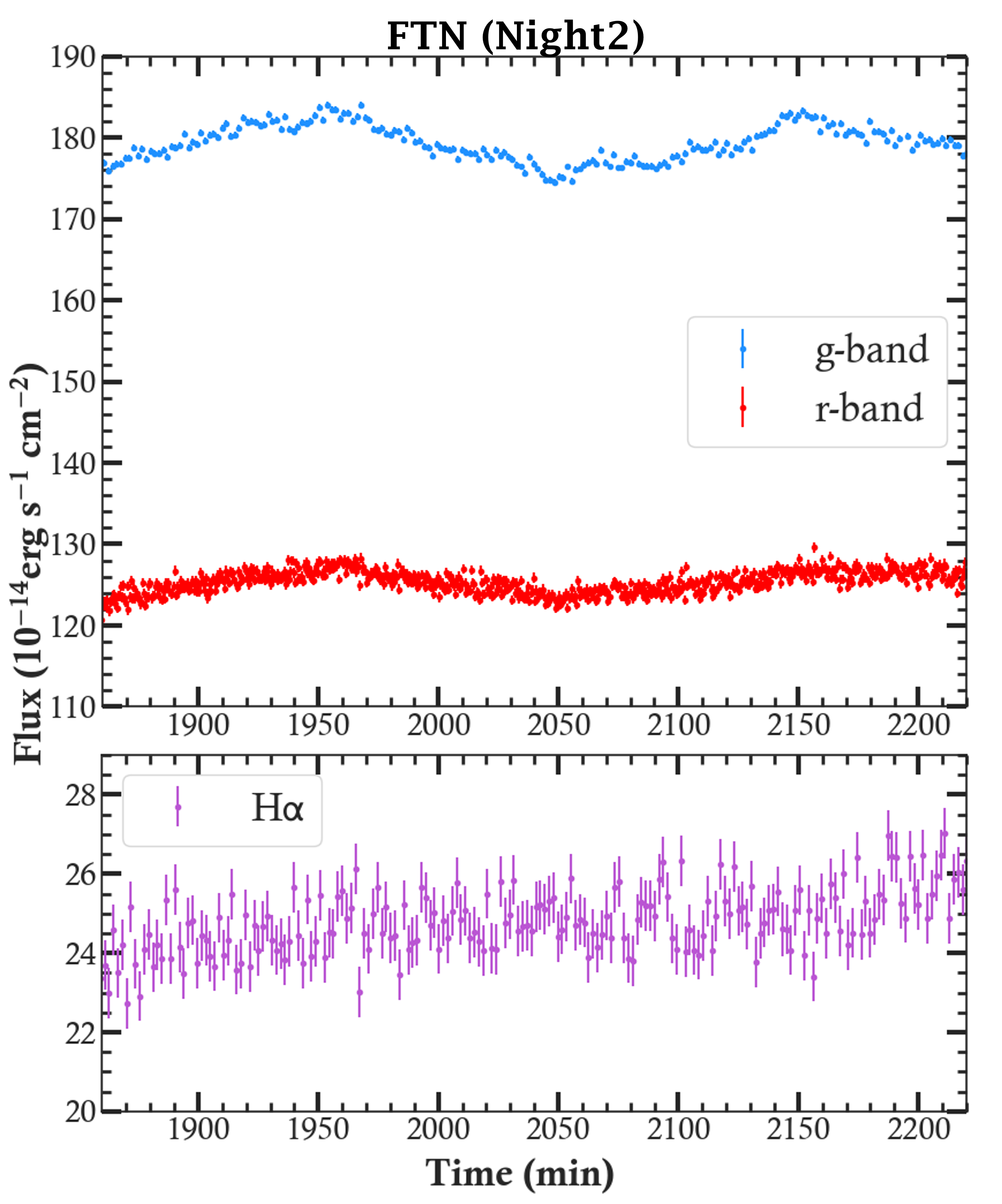}
   \includegraphics[width=60mm]{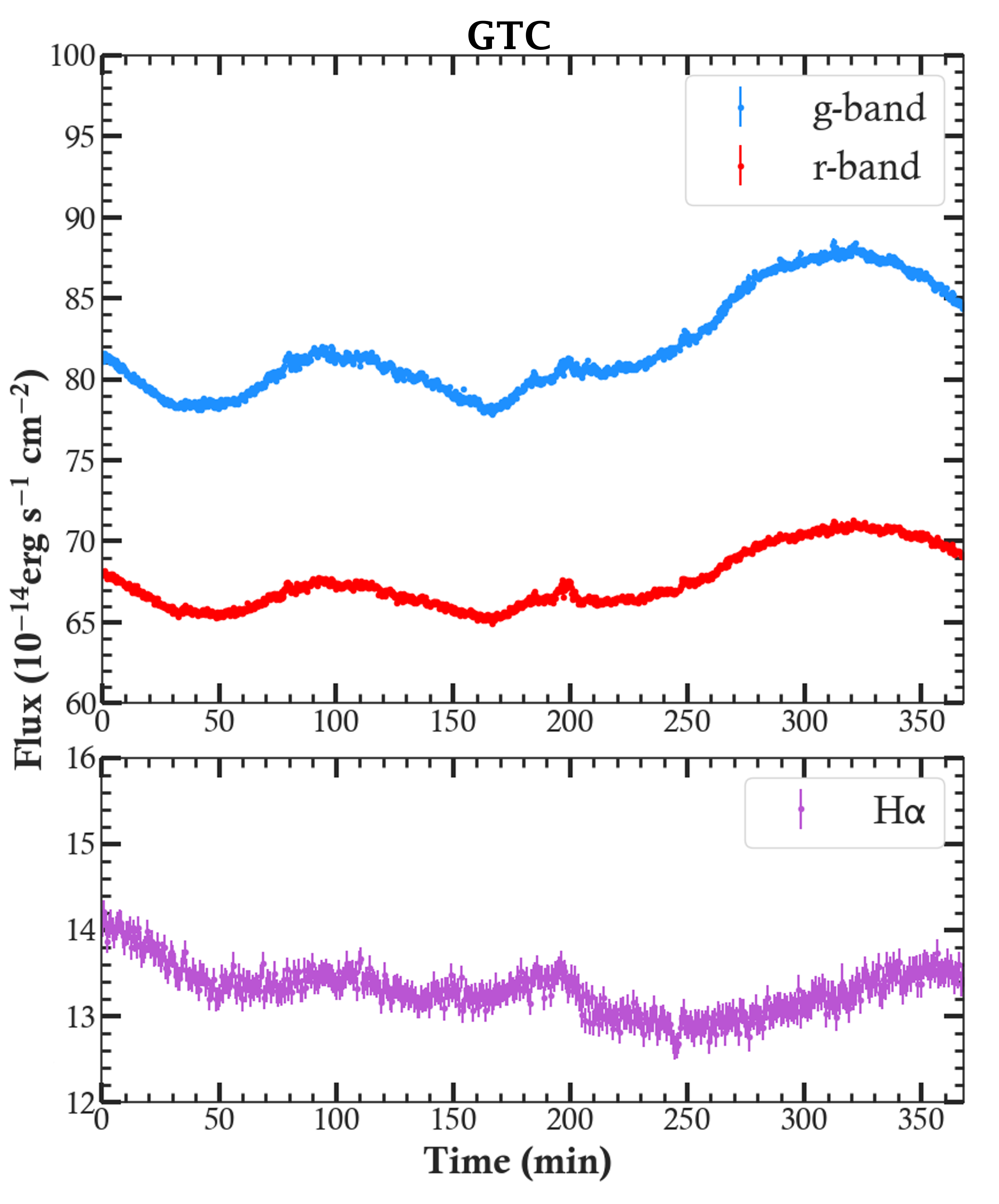}
	  \caption{\label{lc}{\small The light curves of NGC 4395. The upper two panels are the data of FTN for the first and second night respectively, where the starting time is MJD - 59695. The lower panel shows the data from GTC, where the starting time is MJD - 58224.8737808. In the three panels, the g-band (blue) and r-band (red) light curves are sourced from \citet{M} and \citet{McHardy+etal+2023}. The purple lines represent the extracted H$\rm{\alpha}$ light curves using Eq.~(\ref{eq3}).}}
\end{figure}

Building upon the ICCF-Cut method outlined in Sec.~\ref{2.1}, we proceed to extract the broad H$\rm{\alpha}$ light curve out of the r-band photometric data. We first take the FTN g-band and r-band light curves of NGC 4395 in \citet{M} (see Figure~\ref{lc}). The light curves are minute-cadance so that they are able to reveal the small time lag of NGC 4395. In their study, \citet{M} determined that the time lag between g and r-bands is about $5\sim8$ minutes. However, since the r-band encompasses the component of the H$\rm{\alpha}$ emission line, this reported time lag may overestimate the actual time lag reflecting the accretion disk's size. The exact time lag between the g-band and the r-band's continuum component is uncertain, while it will be no more than 5 to 8 minutes. This duration is markedly shorter than the H$\rm{\alpha}$ lag, which is about 1 hour on the order of magnitude \citep{Woo2019}. Therefore, we disregard this minor continuum lag and postulate that the continuum component of the r-band is directly proportional to the g-band light curve. We now employ Eq.~(\ref{eq3}) to extract the broad H$\rm{\alpha}$ light curve. Later, in Sec.~\ref{5.1}, we will revisit this topic to discuss the implications of the continuum lag.

The narrow emission line component remains constant over short timescales. Consequently we first remove it from the r-band. Following Sec.~\ref{3.1}, the narrow emission lines make up $18\%$ of the r-band flux in total. Thus, we subtract a constant, which equals $18\%$ of the mean flux of the light curve, from the r-band. Subsequently, we use Eq.~(\ref{eq3}) to obtain the H$\rm{\alpha}$ broad line component, where $\alpha$ is $1 - 14\%/(1-18\%) = 83 \%$. The error of the extracted H$\rm{\alpha}$ light curves is also determined using Eq.~(\ref{eq3}), which orients from the errors of g and r-band light curves. Additionally, we apply the GTC g-band and r-band light curves of NGC 4395 reported in \citet{McHardy+etal+2023}. The time lag between g and r-bands is $3\sim4$ minutes and therefore can be ignored. We follow the above procedure to extract the light curve of the broad H$\rm{\alpha}$ line as well. The extracted H$\rm{\alpha}$ light curves, along with the original data, are shown in Figure~\ref{lc}.

\section{Results}
\label{sect:result}

In this section we calculate the time lag between the H$\rm{\alpha}$ and the continuum emissions using the light curves presented in Sec.~\ref{3.2}. This time lag is indicative of the BLR size. Subsequently, we will estimate the BH mass. The results from the ICCF-Cut, JAVELIN and $\rm{\chi^2}$ methods are presented below individually.

\subsection{ICCF-Cut Result\label{4.1}}

\begin{figure}
\centering
   \includegraphics[width=90mm]{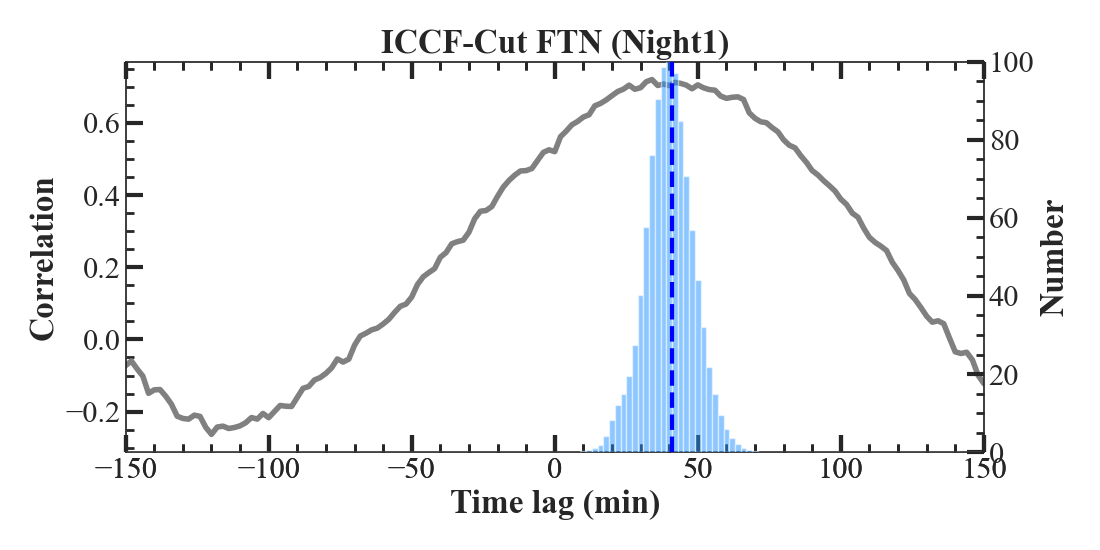}
   \includegraphics[width=90mm]{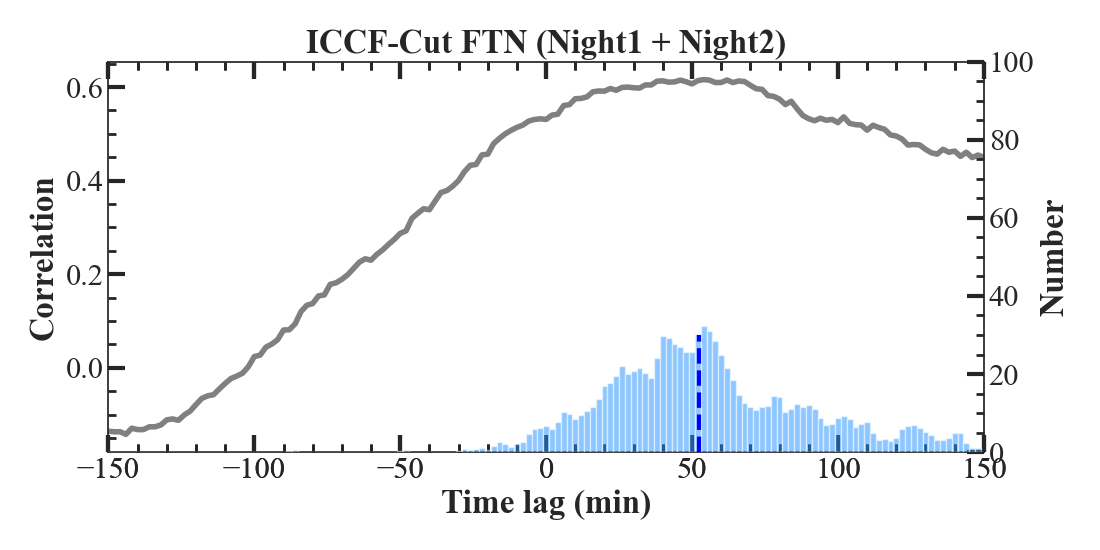}
   \includegraphics[width=90mm]{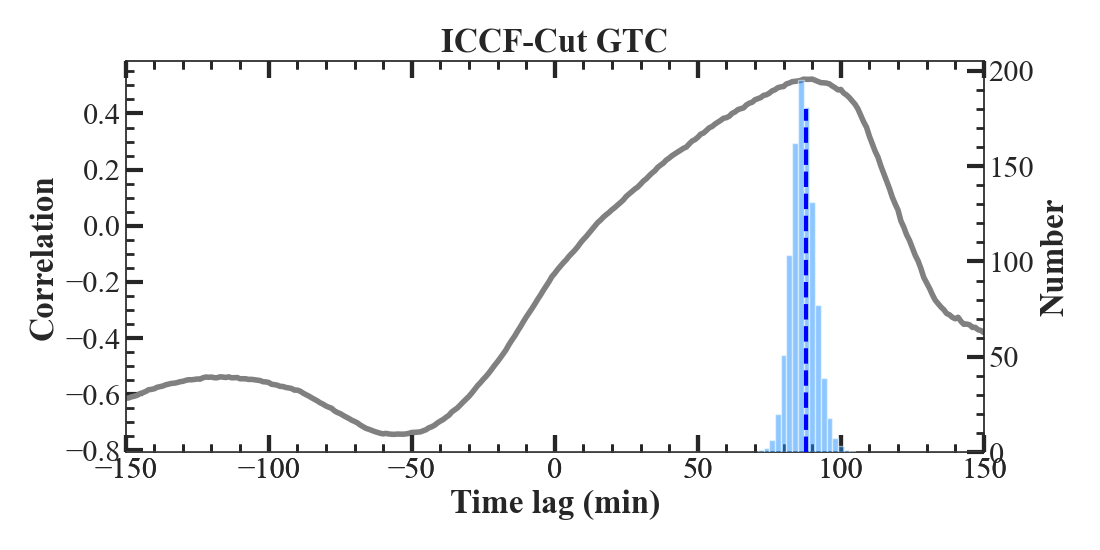}
       \caption{\label{iccfcut}{\small The lag distributions produced by the ICCF-Cut method. The three panels from top to bottom present the broad H$\rm{\alpha}$ line lag distribution derived from the data of: 1. the first night of FTN; 2. the two nights in combination of FTN; 3. the GTC. In three panels, the gray line is the CCF result, and the histogram below is the lag distribution from the FR/RSS method. The dashed lines denote the centroid of the time lag.}}
   
   \end{figure}

We calculate the cross correlation functions (CCFs) of the H$\rm{\alpha}$ and g-band light curves, where we set the lag range from -150 to 150 minutes, with a sampling interval of 1 minute. To estimate the error of the time lag, we employ the Flux Randomization/Random Subset Selection (FR/RSS) method \citep{Peterson1998}, which consists 1000 resampling iterations. Regarding the FTN light curves from \citet{M}, observational data span two nights. We make calculations on the combined light curves of both nights, as well as exclusively on the data from the first night\footnote{The data of the second night exhibits reduced reliability owing to a lower fractional variability amplitude \citep{M}.}. For the GTC light curves from \citet{McHardy+etal+2023}, the time lag is calculated using the full 6-hour data\footnote{The data of the first 8000s is more reliable according to \citet{McHardy+etal+2023}. However, this duration is inadequate to estimate the H$\rm{\alpha}$ lag, whose timescale is about 1 hour. We adopt the whole 6-hour light curves in calculating the time lag, while we need be cautious about the data in the period of 10000 $\sim$ 14000s, which may be affected by the loss of tracking and motion of the observing objects, as noted in \citet{McHardy+etal+2023}.}. These results are all depicted in Figure~\ref{iccfcut}. 

\begin{figure}
  \centering
   \includegraphics[width=70mm]{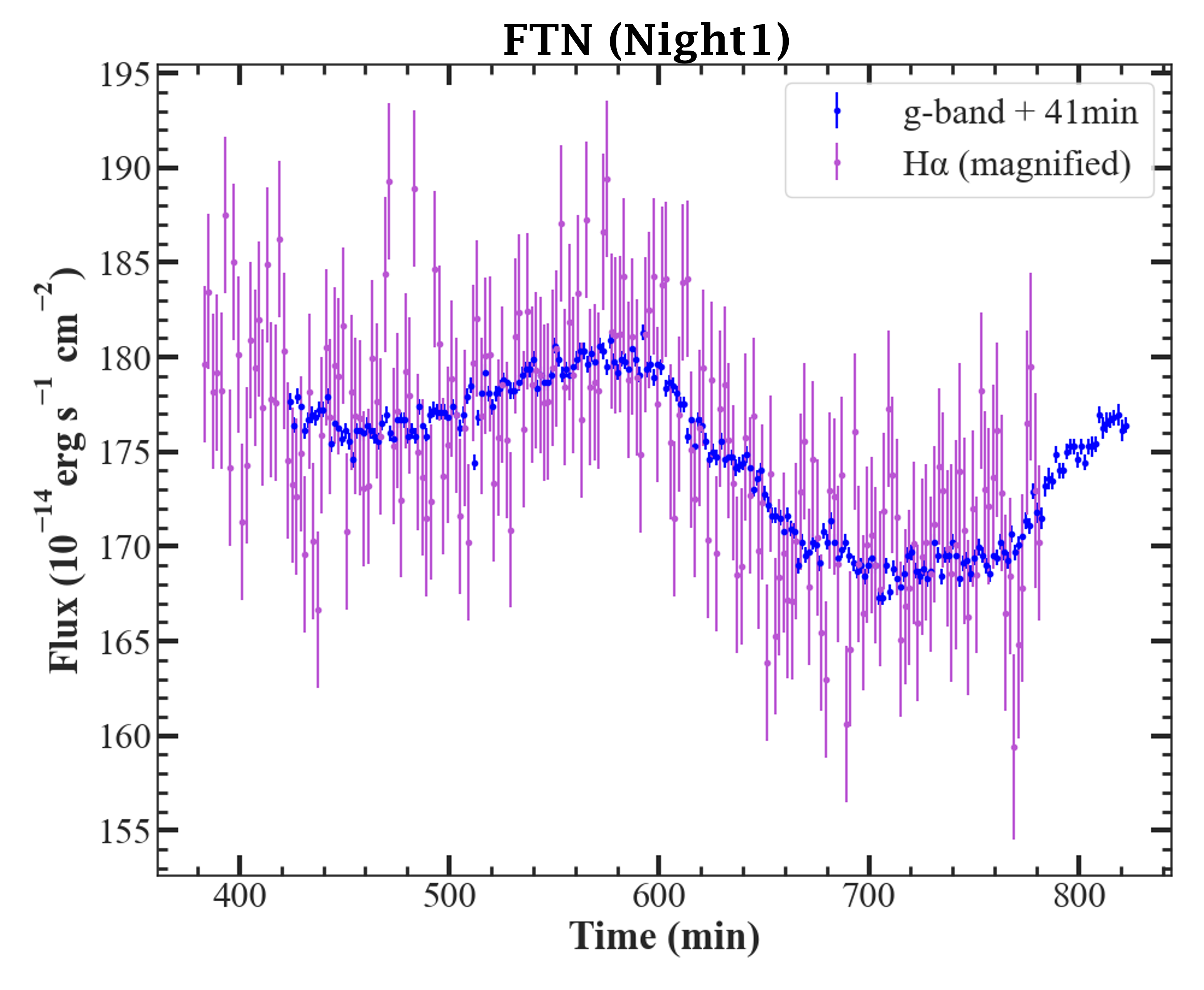}
   \includegraphics[width=70mm]{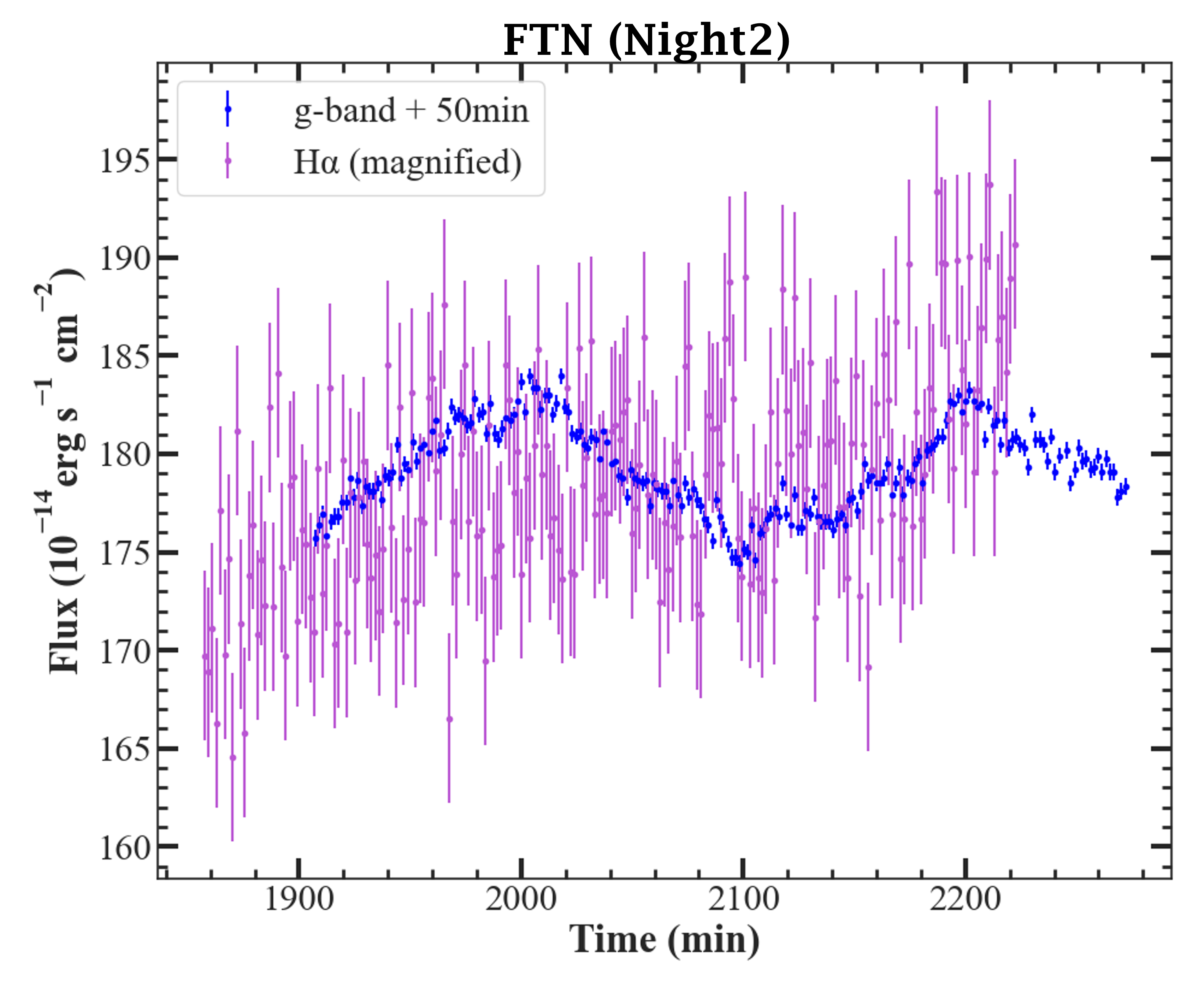}
   \includegraphics[width=70mm]{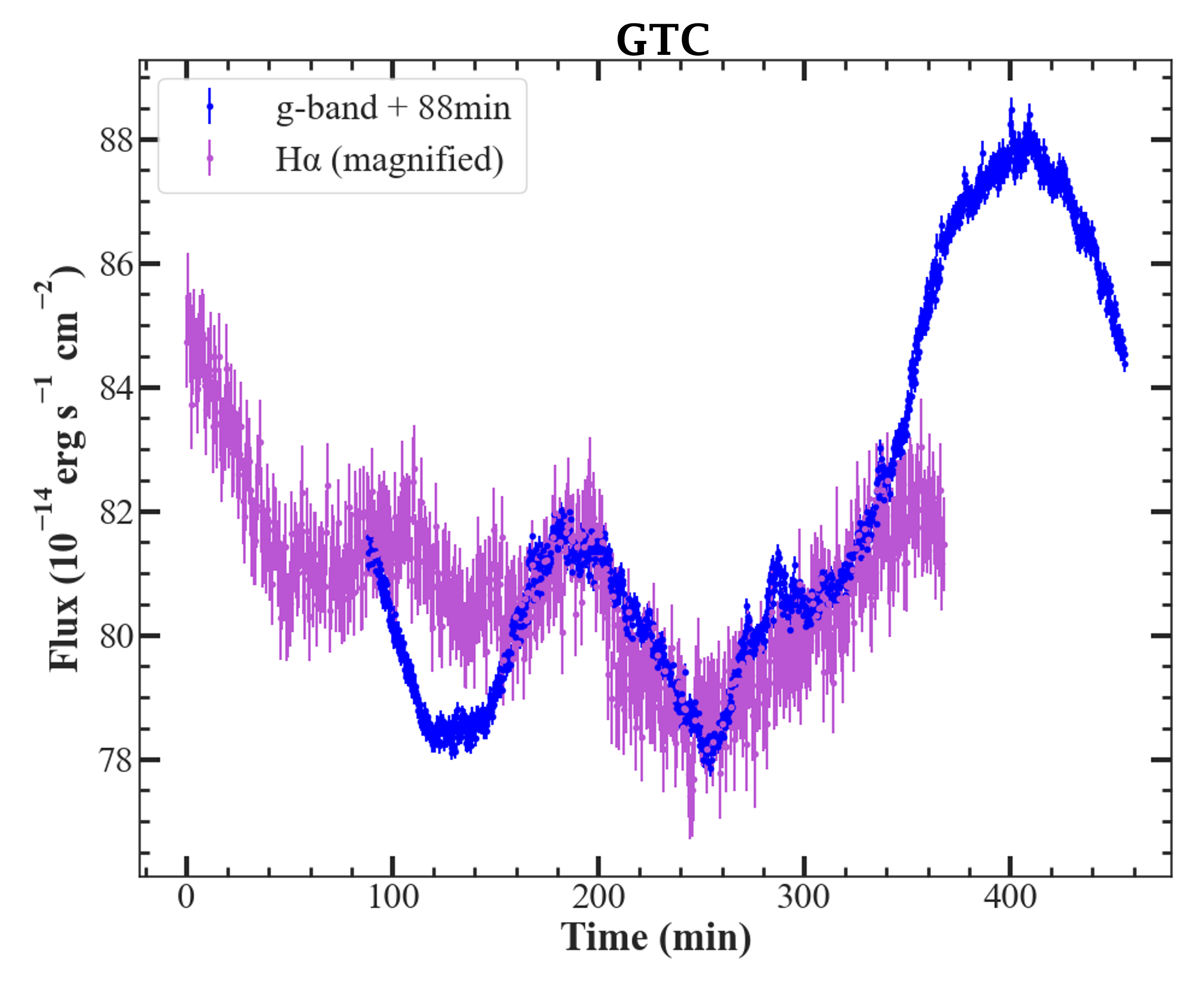}
   \caption{\label{move}{\small The comparison of the extracted H$\rm{\alpha}$ light curves and the lagged g-band light curves. Three panels from top to the bottom are for the light curves from two nights of FTN and from GTC, respectively. Since the flux of H$\rm{\alpha}$ is lower than the continuum flux, we have applied a linear amplification to the H$\rm{\alpha}$ light curves (presented by the purple lines) in order to make the comparison clearer. The blue lines are the lagged g-band light curves, which are shifted by 41 minutes, 50 minutes and 88 minutes, respectively.}}
   
   \end{figure}

For the FTN data, the H$\rm{\alpha}$ time lags determined via the ICCF-Cut method are as follows: for Night 1 alone, the lag is $41^{+8}_{-8}$ minutes with a maximum correlation coefficient $r_{\rm max}$ of $0.72$; for Night 1 + Night 2: the lag is $52^{+39}_{-29}$ minutes with $r_{\rm max}=0.61$. The large uncertainty in the result of Night 1 + Night 2 is probably attributed to the less reliable data from Night 2. For the GTC data, the H$\rm{\alpha}$ time lag is $88^{+4}_{-4}$ minutes with $r_{\rm max}=0.53$ (all the time lags are given at $\rm{1\sigma}$ credible level (CL)). By contrast, we also employ the CCF-ACF method \citep{2012ApJ...747...62C} to compute the H$\rm{\alpha}$ time lag, which yields a time lag of about $44$ minutes with $r_{\rm max}=0.09$ for the FTN Night 1 light curves, and a time lag of around $140$ minutes with $r_{\rm max}=0.25$ for the GTC light curves. Consequently, the results given by ICCF-Cut method are more robust and reliable.

To corroborate the validity of the acquired time lags, we proceed to juxtapose the extracted H$\rm{\alpha}$ light curves with the lagged g-band light curves. Our primary objective is to ascertain whether the g-band light curves, if shifted by the obtained time lag, are capable of aligning with the extracted H$\rm{\alpha}$ light curves. An illustrative representation of this process is provided in Figure~\ref{move}. The congruence between the extracted H$\rm{\alpha}$ light curves and the lagged g-band light curves, as observed in the figure, supports the time lag deduced from the ICCF-Cut method.

\subsection{JAVELIN Result\label{4.2}}

\begin{figure}
 \centering 
   \includegraphics[width=90mm]{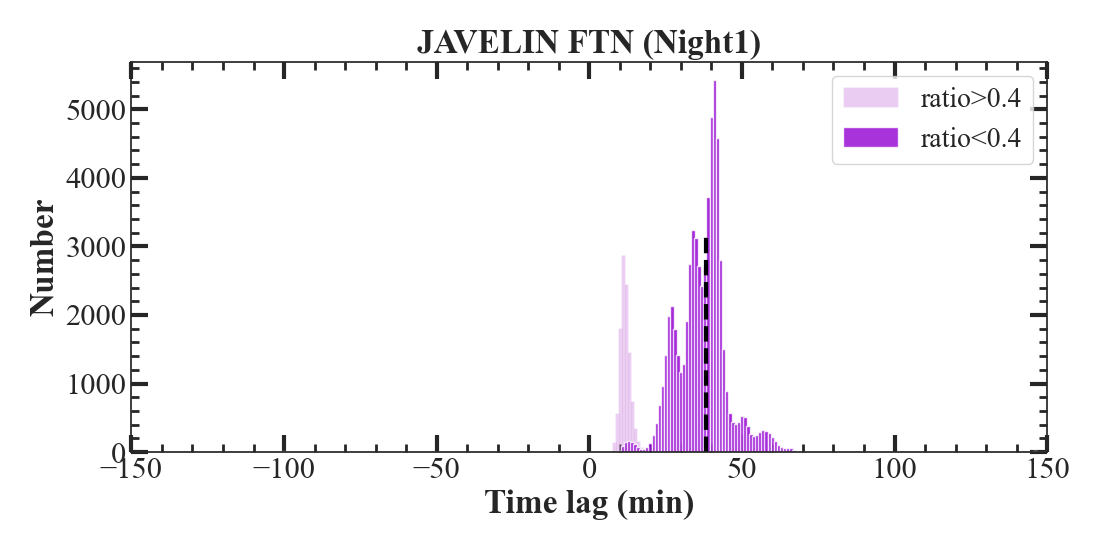}
   \includegraphics[width=90mm]{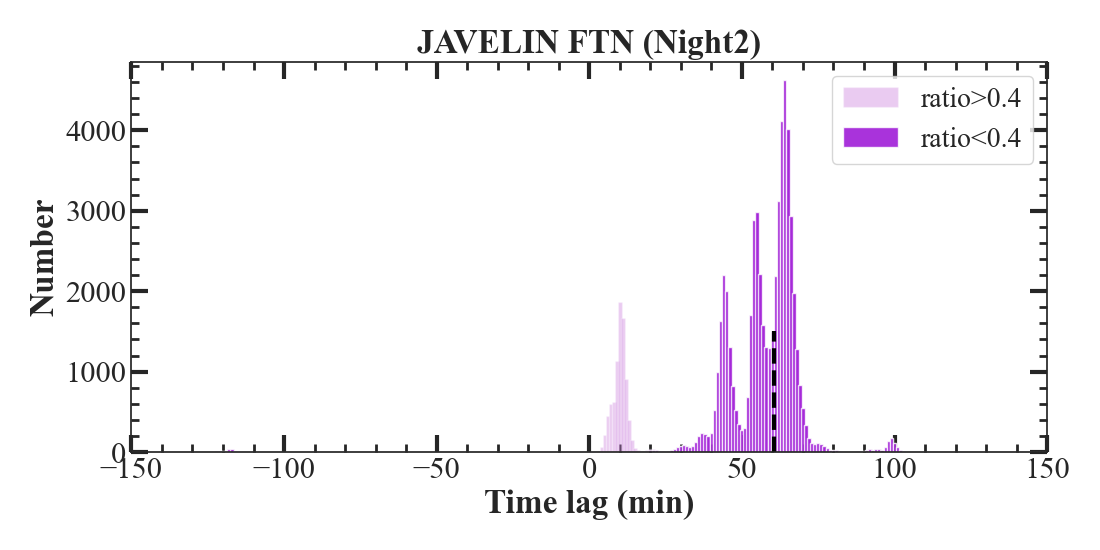}
   \includegraphics[width=90mm]{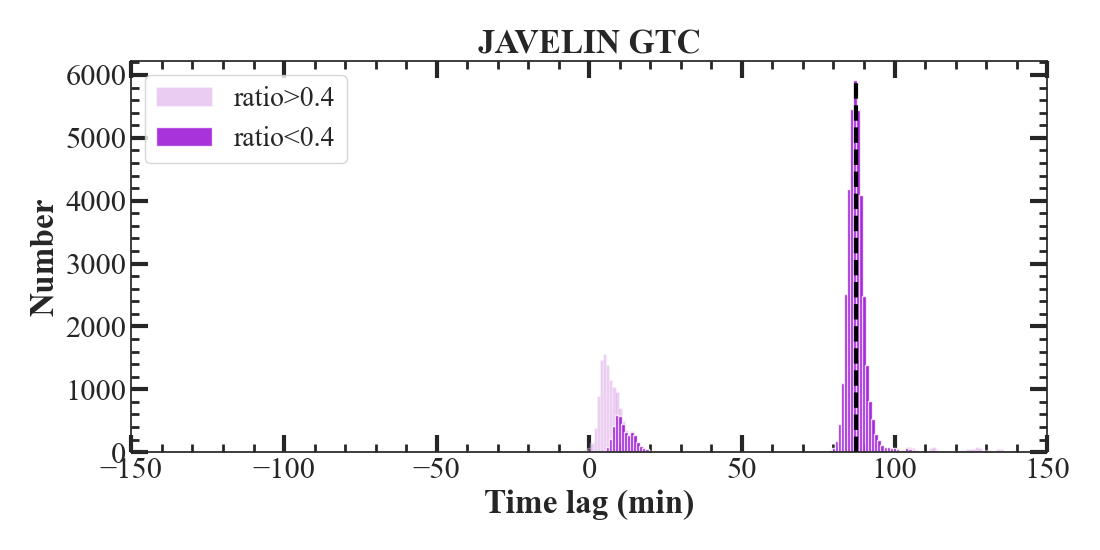}
   \caption{\label{javelin} {\small The results of the H$\rm{\alpha}$ lag distribution from the JAVELIN Pmap model. Three panels from top to the bottom correspond to the light curves of two nights from FTN and from GTC , respectively. The purple regions represent the histograms of H$\rm{\alpha}$ time lag. The data points with ratio $\textgreater 0.4$ are not physical (shown in light purple) and should be neglected.}}   
   \end{figure}

\begin{figure}
 \centering 
   \includegraphics[width=75mm]{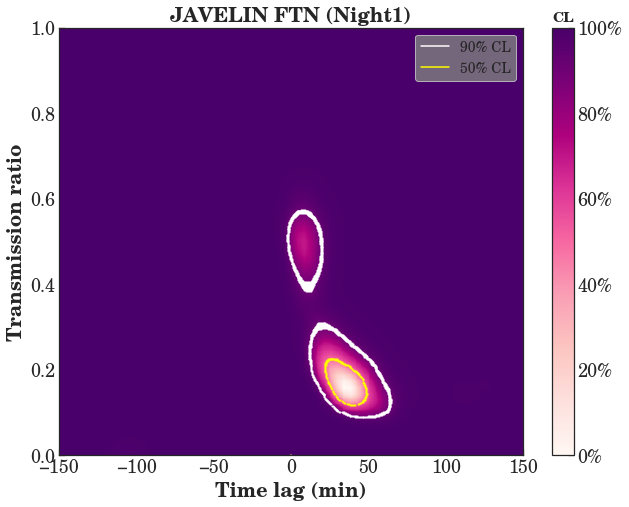}
   \includegraphics[width=75mm]{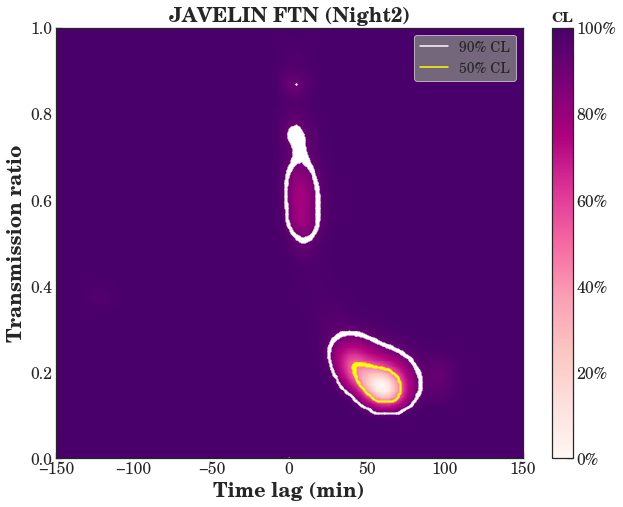}
   \includegraphics[width=75mm]{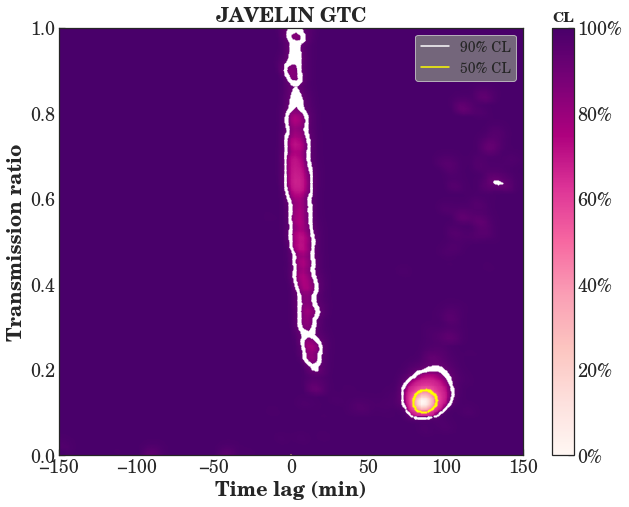}
   \caption{\label{ratio}{\small The distributions of H$\rm{\alpha}$ transfer function ratio in relation to the time lag. The color-coded regions represent varying credible levels (CLs) in the lag-ratio space. The regions with lighter color indicate higher probabilities. The boundaries of the $90\%$ CL and $50\%$ CL regions are marked by white and yellow lines, respectively.}}
   \end{figure}

To further evaluate the reliability of the computed time lag, we now employ the Pmap model of JAVELIN to determine the H$\rm{\alpha}$ time lag of NGC 4395. Subsequently, these results will be compared with those obtained from the ICCF-Cut method. In this model, the input r-band light curve is regarded as the superposition of the continuum and a lagged H$\rm{\alpha}$ component. The time lag between them is then determined through 90000 MCMC processes. For the light curves provided by \citet{M} and \citet{McHardy+etal+2023}, the time lag distributions are depicted in Figure~\ref{javelin}\footnote{When using JAVELIN, it is inappropriate to combine the light curves of two nights, for JAVELIN is based on the DRW model and this may introduce unphysical variability to the light curves within the gap between the two nights.}. In Figure~\ref{ratio}, we also present the distribution of the transfer function ratio (hereafter ratio) for the H$\rm{\alpha}$ component, which reflects the flux ratio of the broad H$\rm{\alpha}$ in the r-band. From Figure~\ref{ratio}, it can be found that the distributions of the lag-ratio exhibit  ``double peaks". The peak located in the middle is artificial, whose time lag is near zero while the ratio is very high. This peak is unphysical and is only mathematically valid\footnote{It is straightforward to deduce that if the ratio is $100\%$, implying that the r-band is considered as a pure line component, then the JAVELIN Pmap is reduced to Rmap and the obtained time lag is just the lag between g and r-bands, which is several minutes \citep{M}. Therefore we have the peak with high ratio and small time lag in Figure~\ref{ratio}, which is merely an artefact of the JAVELIN Pmap. Consequently, we need to restrict the ratio based on the actual spectrum.}. For the samples with transfer function ratio $\textgreater$ $0.4$, they are not physically feasible since the observed broad H$\rm{\alpha}$ flux ratio is only $14\%$. The value of $0.4$ is set to eliminate most of the contribution from the middle peak, while retaining the contribution from the other peak. When calculating the ranges of the time lag, we have excluded these high ratio parts.

For the FTN data, the H$\rm{\alpha}$ time lags are $38^{+6}_{-12}$ minutes and $58^{+7}_{-15}$ minutes for the two nights respectively. For the GTC data, the H$\rm{\alpha}$ time lag is $87^{+4}_{-5}$ minutes (all at $\rm{1\sigma}$ CL). Moreover, the transfer function ratios of H$\rm{\alpha}$ are all centered around $15\%$, which aligns with the observed H$\rm{\alpha}$ flux ratio ($14\%$). This consistency adds to the credibility of both the ICCF-Cut method and the JAVELIN method. The H$\rm{\alpha}$ time lags as determined by JAVELIN also agree with the results from the ICCF-Cut method.

\subsection{$\rm{\chi^2}$ Result\label{4.3}}

\begin{figure}
  \centering 
   \includegraphics[width=90mm]{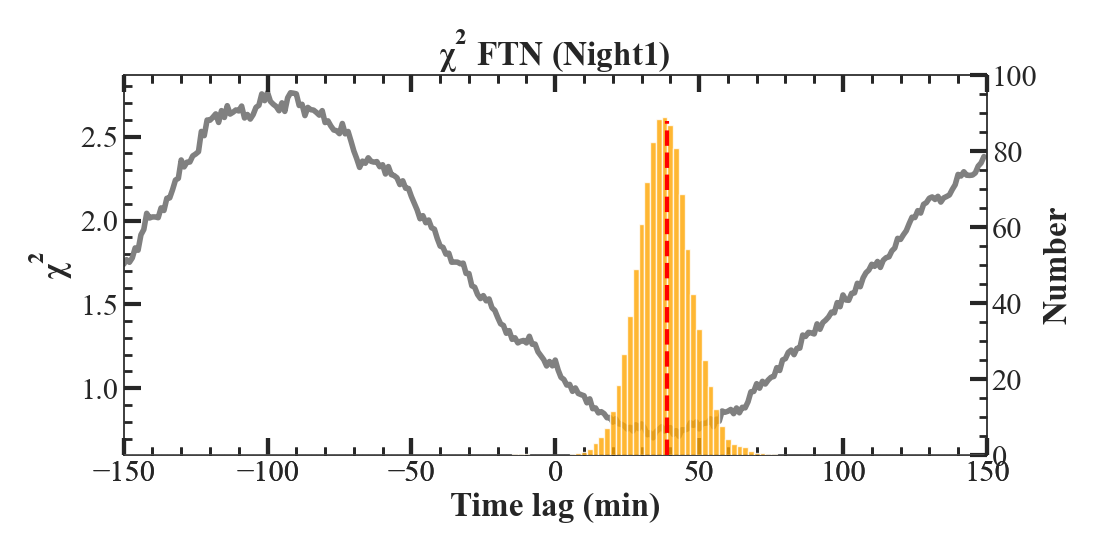}
   \includegraphics[width=90mm]{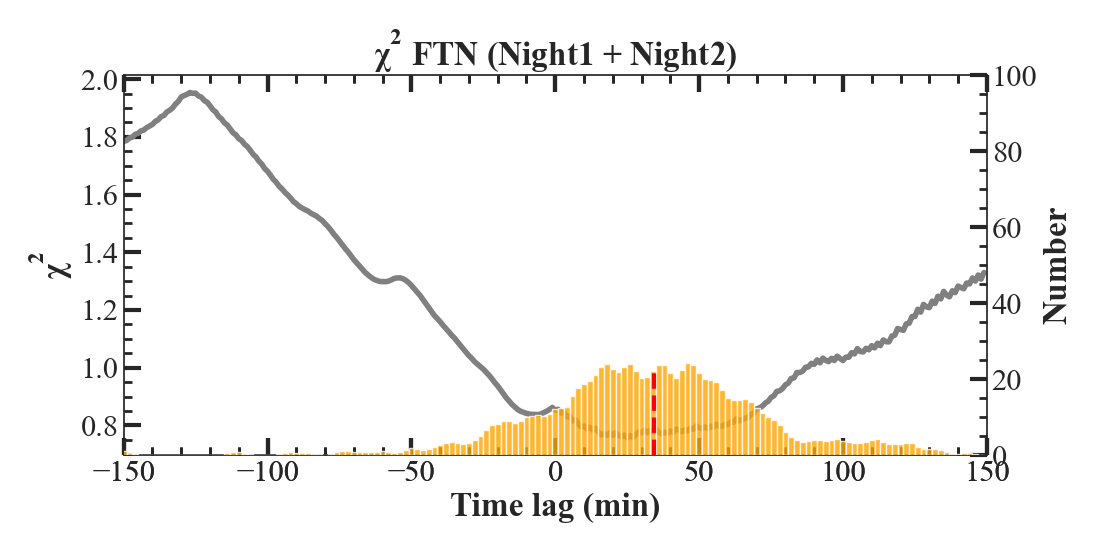}
   \includegraphics[width=90mm]{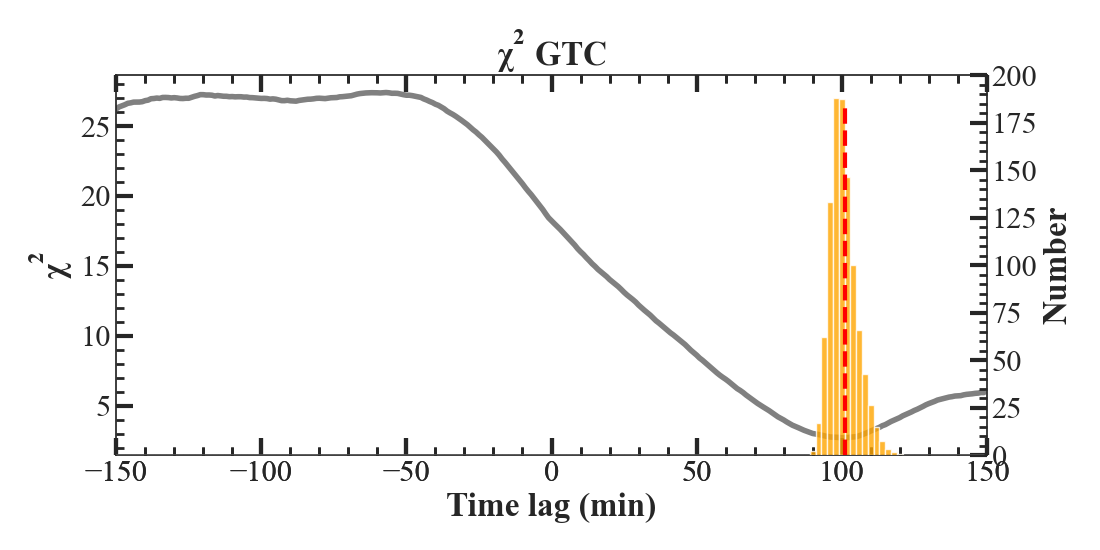}
   \caption{\label{chi2}{ \small The results of the $\rm{\chi^2}$ method. All the notations are analogous to those in Figure~\ref{iccfcut}, except that the gray lines now represent the $\rm{\chi^2}$ as a function of time lag.}}

   \end{figure}

In this part we employ the third method : the $\rm{\chi^2}$ method to calculate the H$\rm{\alpha}$ time lags. As outlined in Sec.~\ref{2.3}, the $\rm{\chi^2}$ method focuses on the uncertainties in the two light curves. We compute the $\rm{\chi^2}$ between the g-band light curves and the extracted H$\rm{\alpha}$ light curves as a function of the time delay, whose minimum point reflects the most probable lag. We conducted 1,000 iterations of FR/RSS to analyse the uncertainties of the time lags, for the data of Night 1, Night 1 + Night 2 from FTN and from GTC separately. The results are shown in Figure~\ref{chi2}.

For the FTN data, the H$\rm{\alpha}$ time lags are $39^{+8}_{-8}$ minutes for Night 1 and $34^{+34}_{-35}$ minutes for the combined data of Night 1 + Night 2. For the GTC data, the H$\rm{\alpha}$ time lag is $101^{+4}_{-3}$ minutes (all at $\rm{1\sigma}$ CL). These results are in alignment with those obtained from the ICCF-Cut and JAVELIN methods.

In general, the results of three methods are listed in Table~\ref{Tab2}, which exhibit considerable consistency. For the FTN observations, the data in Night 2 is not as reliable as that from Night 1, so we need to be cautious about the results for Night 1 + Night 2. Across the three methods, the H$\rm{\alpha}$ lag is about $40$ minutes for the first night from FTN and $90$ minutes from GTC. The latter value is comparable with the $83$ min result given by \citet{Woo2019} using narrow-band photometry.

\subsection{R-L Relation\label{4.4}}

With the H$\rm{\alpha}$ lag measured in the range of $\rm{40\sim90}$ minutes, we investigate the relationship between its BLR size and luminosity, specifically the $R_{\rm BLR}-L_{\rm AGN}$ relation \citep{RL}. We calculate the average g-band and r-band flux for the light curves in \citet{M} and \citet{McHardy+etal+2023} separately, and then estimate the flux at $\rm{5100 \mathring{A}}$ from the average flux density in the g and r-bands. The $5100\mathring{A}$ luminosity $\lambda L_{\rm \lambda,5100\mathring{A}}$ for the two light curves are calculated as $\rm{1.2\times 10^{40}erg \ s^{-1}}$ and $\rm{1.1\times 10^{40}erg \ s^{-1}}$ respectively, which are consistent with the measurement in \citet{Cho2020}, where $\lambda L_{\rm \lambda,5100\mathring{A}}=1.0\times 10^{40}\rm erg \ s^{-1}$.

\begin{figure}
   \centering
   \includegraphics[width=8cm, angle=0]{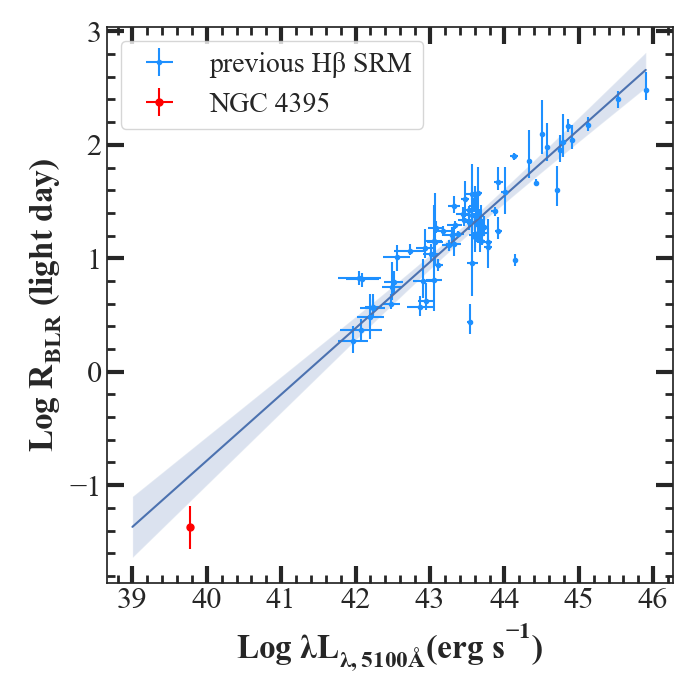}
   \caption{\label{fig-RL}{\small The radius-luminosity relation for NGC 4395 (red dot), together with samples from \citet{RL2} (blue dots). The straight line shows the best fit for the samples in \citet{RL2}. The regions within the error margin ($\sigma=0.2$dex) are marked in light blue.}}

 \end{figure}

This luminosity, however, includes contributions from its host galaxy and the NSC. The result from \citet{Cho2020} indicates that the total luminosity of the two components is approximately $\rm{5\times 10^{39}erg \ s^{-1}}$. We assume this to be invariant with time, so the luminosity of the AGN is about $\lambda L_{\rm \lambda,5100\mathring{A}}=6\times 10^{39}\rm erg \ s^{-1}$. This value is similar to the $\lambda L_{\rm \lambda,5100\mathring{A}}=5.75\times 10^{39}\rm erg \ s^{-1}$ reported in \citet{Cho2020}. We then plot our result in the $R_{\rm BLR}-L_{\rm AGN}$ space. Previous studies suggest an estimation that
\begin{equation}
\log (R_{\mathrm{BLR}} / \mathrm{1 lt-day} )=K+\alpha \log \left(\lambda L_{\lambda,5100\mathring{A}} / 10^{44} \mathrm{erg} \mathrm{s}^{-1}\right),
\end{equation}
where ${K=1.527_{-0.031}^{+0.031}, \alpha=0.533_{-0.033}^{+0.035}}$ according to some previous H$\rm{\beta}$ RM campaigns \citep{RL2}.
This relation is based on AGN samples with luminosity $\lambda L_{\rm \lambda,5100\mathring{A}} > \rm 10^{42}erg \ s^{-1}$. For a long time, the relationship was missing for low luminosity samples. Our result of NGC 4395 is shown in Figure~\ref{fig-RL}, along with the previous samples in \citet{RL2}. It's inspiring to see that NGC 4395 matches the extrapolated R-L relation well, proving that the R-L relation holds approximately at the low luminosity end as well.   

\subsection{Black Hole Mass Estimation\label{4.5}}

We apply Eq.~(\ref{eq1}) to estimate the central black hole mass of NGC 4395. The FWHM of the broad H$\rm{\alpha}$ line is $\rm{740 \pm 3}$ $\rm{km \ s^{-1}}$. We adopt the scale factor $f$ from \citet{Woo2015}: $f = \rm{1.1\pm 0.3}$. It is worth noting that the FWHM of the broad H$\rm{\alpha}$, derived from the single-epoch spectrum discussed in Sec.~\ref{3.1}, is not synchronous with the photometric data in Sec.~\ref{3.2}. Since the luminosity of the AGN varies over time, the FWHM of the broad H$\rm{\alpha}$ line that we obtained may be higher or lower than its average value, leading to potential inaccuracies in the determination of the BH mass. By combining this value with the H$\rm{\alpha}$ lag between $\rm{40\sim90}$ minutes, the estimated BH mass of NGC 4395 is $\rm M_{BH} = (8 \pm 4) \times 10^3\, M_{\odot}$. If utilizing the FWHM $\sim 1000$ $\rm{km \ s^{-1}}$ measured in April 2017 \citep{Woo2019}, the BH mass is estimated to be $\rm M_{BH} = (12 \pm 5) \times 10^3\, M_{\odot}$. The primary source of uncertainty in these outcomes is the divergent H$\rm{\alpha}$ line lag obtained from the two epochs. Our result aligns with \citet{Woo2019}, who reported a black hole mass of $\rm (9.1 \pm 1.6)\times 10^3\,M_{\odot}$ from the narrow-band photometry.

\begin{table}
\begin{center}
\caption[]{H$\rm{\alpha}$ time lag (min) obtained from the three methods}\label{Tab2}

\begin{tabular}{cccc}
\hline
Method          & FTN Night 1  & FTN Night 1 + Night 2 & GTC \\ \hline
ICCF-Cut          & $41^{+8}_{-8}$ &  $52^{+39}_{-29}$  & $88^{+4}_{-4}$          \\ 
JAVELIN & $38^{+6}_{-12}$    &   $58^{+7}_{-15}$   & $87^{+4}_{-3}$        \\ 
$\rm{\chi^2}$ & $39^{+8}_{-8}$  &   $34^{+34}_{-35}$   & $101^{+4}_{-3}$        \\ 
\hline
\end{tabular}
\end{center}
\end{table}

\section{Discussion}
\label{sect:discussion}
\subsection{Influence of the Continuum Lag\label{5.1}}
In Sec.~\ref{3.2}, the time lag between the g and r-band continuum components was initially disregarded during the extraction of the H$\rm{\alpha}$ light curves. This may have overlooked the potential impact of such time lag on our analysis. To address this, we introduce a possible continuum lag and translate the g-band light curves by the time lag. Then the shifted g-band light curve is regarded as part of the r-band continuum component. This procedure is demonstrated using the GTC light curves, assuming a 4-minute time delay between the g and r-band continuum components. Following this adjustment, we use ICCF-Cut to calculate the H$\rm{\alpha}$ time lag. The results indicate that the central H$\rm{\alpha}$ time lag remains at $88$ minutes, but with a reduced correlation factor from $0.5$ to $0.25$.

The inferiority of this outcome, when compared to the previous result, can be interpreted as follows. It has been reported that the lag between the g and r-bands is $5\sim8$ minutes according to \citet{M} and $3\sim4$ minutes according to \citet{McHardy+etal+2023}. However, it is imperative to recognize that this lag may not be exactly the continuum lag of the accretion disk, because the broad H$\rm{\alpha}$ line makes up $15\%$ of the total flux, whose contribution to the g-r lag cannot be neglected. Given that our analysis is confined to photometric data, the genuine H$\rm{\alpha}$ lag and the continuum lag are probably degenerate. Anyway, our test suggests that the H$\rm{\alpha}$ lag result remains relatively unaffected provided that the lag between the g and r-bands is sufficiently minor.

\subsection{Influences of the NSC and Host Galaxy\label{5.2}}

As mentioned in Sec.~\ref{3.2}, the photometric data of NGC 4395 includes contributions from a NSC and the host galaxy, which could potentially influence the results derived from the ICCF-Cut method. In the following, we discuss their effects from two perspectives, utilizing the light curves obtained from GTC \citep{McHardy+etal+2023} as an illustration.

Firstly, the justification for excluding the host galaxy and NSC components is rooted in the ICCF-Cut method's fundamental premise, where flux variations are proportional to the mean flux across all bands. In an ideal scenario, the light curves emanating from the accretion disk would follow this principle, i.e. one would anticipate that the bands with a higher average flux will exhibit greater flux variability. However, the flux contributions from the host galaxy and the NSC remain constant over time. Consequently, the total flux from broad-band observation may not maintain a proportional relationship with its variation, since the constant component could be arbitrarily large. As a result, we usually need to cut out the contributions from the host galaxy and the NSC to ensure that the residual flux satisfies the expected proportional relation.

As observed in Figure~\ref{lc}, the flux variations of the g and r-bands exhibit a ratio of approximately $\rm{3:2}$, which is consistent with the ratio of their average fluxes after accounting for the removal of emission line contributions in the r-band. This consistency suggests that the influence of the host galaxy and the NSC can be disregarded in this context, as their flux contributions should also satisfy this proportional relation.

Secondly, we performed a calculation to determine the exact contribution of the host galaxy and the NSC in the g and r-bands for NGC 4395. The $\rm{5100\mathring{A}}$ luminosity for its host galaxy and the NSC were measured as $\rm{1.6\times 10^{39}erg \ s^{-1}}$ and $\rm{3.6\times 10^{39}erg \ s^{-1}}$, respectively, as reported by \citet{Cho2020}. The SEDs for both are presented in \citet{Carson2015}. Our analysis revealed that their contribution is about about $\rm{2.6\times 10^{-13}erg \ s^{-1} \ cm^{-2}}$ in the g-band and $\rm{2.4\times 10^{-13}erg \ s^{-1} \ cm^{-2}}$ (including $\rm{8\times 10^{-14}erg \ s^{-1} \ cm^{-2}}$ from the narrow emission lines) in the r-band, for the GTC data utilized. After excising these contributions from the light curves, which results in an improvement in the broad H$\rm{\alpha}$ flux ratio in the r-band to $25\%$, the ICCF-Cut method was reapplied. The measured time lag for broad H$\rm{\alpha}$ is $85^{+4}_{-4}$ minutes, consistent with the previously reported $88^{+4}_{-4}$ minutes in Sec.~\ref{4.1}. Consequently, it appears justifiable to discount the impact of the host galaxy and the NSC, as they do not significantly alter the lag measurement.

\section{Conclusion}
\label{sect:conclusion}
In this paper, we present our work on H$\rm{\alpha}$ reverberation mapping for NGC 4395 using broad-band photometric data. NGC 4395 is a dwarf AGN with rapid flux variations. By analyzing minute-cadence g and r-band light curves sourced from recent researches \citep{M,McHardy+etal+2023}, we calculate its H$\rm{\alpha}$ time lag and central BH mass through a combinative application of the ICCF-Cut, JAVELIN and $\rm{\chi^2}$ methods. The results are useful in both improving our cognition to the AGNs in the low luminosity end and demonstrating the feasibility of using the broad-band photometric data for H$\rm{\alpha}$ reverberation mapping.

We utilize a single-epoch spectrum of NGC 4395 from \citet{Cho2021} to extract the H$\rm{\alpha}$ component from the broad-band light curves. The broad H$\rm{\alpha}$ emission line accounts for $14\%$ of the flux in the r-band. Our calculations of the H$\rm{\alpha}$ time lags exhibit consistency when applied across three distinct methods. For the light curves from \citet{M}, the H$\rm{\alpha}$ time lag is $41^{+8}_{-8}$ minutes from ICCF-Cut method. For the light curves from \citet{McHardy+etal+2023}, the H$\rm{\alpha}$ time lag is $88^{+4}_{-4}$ minutes. Furthermore, the results from the JAVELIN Pmap model suggest that the distribution of the H$\rm{\alpha}$ flux ratio in the r-band is centered at approximately $15\%$, aligning with the spectral observation.

The FWHM of the H$\rm{\alpha}$ broad component for NGC 4395 is calculated to be approximately $740$ $\rm{km \ s^{-1}}$, leading to an estimated central BH mass of $\rm M_{BH} = (8 \pm 4) \times 10^3\, M_{\odot}$. If using the spectrum from \citet{Woo2019}, the BH mass is about $\rm M_{BH} = (12 \pm 5) \times 10^3\, M_{\odot}$. This result corroborates earlier RM studies that employed narrow-band photometry or spectroscopy, thereby reinforcing the hypothesis that NGC 4395 harbors an intermediate-mass BH in its center. Our investigation into the R-L relationship for NGC 4395 confirms that the R-L relation holds for the low luminosity end as well. Looking forward, we anticipate that our H$\rm{\alpha}$ reverberation mapping methods based on the broad-band photometry could be widely applicable to a greater number of AGNs in the era of large multi-epoch, high-cadence photometric surveys.

\section*{Acknowledgements}

We acknowledge the science research grant from the China Manned Space Project with No. CMS-CSST-2021-A06. This work is supported by the National Key R\&D Program of China No. 2022YFF0503401. We are thankful for the support of the National Science Foundation of China (11927804, and 12133001). We thank I. M. McHardy and J.-H. Woo for providing GTC photometric data and Gemini spectroscopic data of NGC 4395. This paper is based on observations made with the Faulkes North Telescope (FTN) and Gran Telescopio Canarias (GTC) telescope. FTN is a part of the Las Cumbres Observatory (LCO) Global Telescope Network which is located at Haleakalā Observatory. GTC is installed at the Spanish Observatorio del Roque de los Muchachos of the Instituto de Astrofísica de Canarias, on the island of La Palma. We thank the staff in planning the FTN MuSCAT3 observations as well as the GTC HiPERCAM observations.

%%%%%%%%%%%%%%%%%%%%%%%%%%%%%%%%%%%%%%%%%%%%%%%%%%

\section*{Data Availability}

The FTN light curves for NGC 4395 used in this paper are published in \citet{M}. The GTC light curves for NGC 4395 used in this paper are available from the link\footnote{\href{https://gtc.sdc.cab.inta-csic.es/gtc/jsp/searchform.jsp}{https://gtc.sdc.cab.inta-csic.es/gtc/jsp/searchform.jsp}} provided by \citet{McHardy+etal+2023}. The Gemini spectrum data used in this paper is shared by \citet{Cho2021}.



\bibliographystyle{mnras}
\bibliography{bibtex} % if your bibtex file is called example.bib

%\begin{thebibliography}{99}
%\end{thebibliography}

%%%%%%%%%%%%%%%%%%%%%%%%%%%%%%%%%%%%%%%%%%%%%%%%%%

%%%%%%%%%%%%%%%%% APPENDICES %%%%%%%%%%%%%%%%%%%%%

\appendix

%%%%%%%%%%%%%%%%%%%%%%%%%%%%%%%%%%%%%%%%%%%%%%%%%%

% Don't change these lines
\bsp	% typesetting comment
\label{lastpage}
\end{document}